\documentclass[12pt]{article}
\usepackage{calrsfs}
\usepackage{indentfirst}
\usepackage{color}
\usepackage{indentfirst}
\usepackage{eucal}
\usepackage{amsmath}
\usepackage{amsfonts}
\usepackage{amssymb}
\usepackage{mathrsfs}
\usepackage{bm}
\usepackage{color}
\usepackage{indentfirst}
\usepackage{eucal}
\usepackage{amsmath}
\usepackage{amssymb}
\usepackage{mathrsfs}
\usepackage{bm}
\usepackage{fancyhdr}
\usepackage[nottoc,notlot,notlof]{tocbibind}
\begin{document}
\begin{center}{ \Large \bf Schr\"{o}dinger equation as a confluent Heun equation}
\vskip 0.4cm {Bartolomeu Donatila Bonorino Figueiredo} \\
{Centro Brasileiro de Pesquisas F\'isicas (CBPF)\\
 Rua Dr. Xavier Sigaud, 150 - 22290-180 - Rio de Janeiro, RJ, Brasil}
\end{center}
%
%
%
%
\begin{abstract}
\noindent
This article deals with two classes of quasi-exactly solvable (QES) 
trigonometric potentials for which the one-dimensional 
Schr\"odinger equation reduces to a confluent Heun
equation (CHE) where the independent variable  takes only finite values. 
Power series for the CHE  are used
to get finite- and infinite-series eigenfunctions. 
Finite series occur only for special sets of parameters and characterize  
the quasi-exact solvability. 
Infinite series occur for all admissible  values of the parameters
(even values involving finite series), and are bounded and convergent in the entire range of the independent variable.
Moreover, throughout the article we examine other QES  trigonometric 
  and hyperbolic  potentials.  In all cases,  for a finite series there is a convergent infinite series. 


\tableofcontents

\end{abstract}
%
%
%

\newpage

\section{Introduction}
 
 We reconsider the 
power series solutions for the confluent Heun equation (CHE) and use them to find systematically finite- and infinite-series 
eigenfunctions  of the Schr\"{o}dinger equation with quasi-exactly solvable (QES) 
trigonometric potentials. 
We examine as well other QES potentials  given by trigonometric 
  and hyperbolic functions.

In fact we are concerned with solutions for the one-dimensional 
stationary Schr\"{o}dinger 
equation  written as  
\begin{eqnarray}
\label{schr}
\frac{d^2\psi(u)}{du^2}+\big[{\cal E}-{\cal V}(u)\big]\psi(u)=0, 
\end{eqnarray}
where ${\cal E}$ takes the place of the energy of the particle
and ${\cal V}(u)$ is a function proportional  to the potential. For ${\cal V}(u)$
we use  two classes of  Ushveridze's trigonometric potentials  
 given by  \cite{ushveridze} 
\begin{eqnarray}
&&\label{ush-1-trigonometrico}
\begin{array}{l} \mathcal{V}_1(u)=
-4\beta^{2}{\sin^{4}u}+ 4\beta\big(\beta+
2\gamma+2\delta+2\ell \big){\sin^{2}u}+
\frac{\left( 4\gamma-1\right)\left(4\gamma-3\right)}
{4\cos^{2}u}+\frac{\left(4 \delta-1\right)
\left(4\delta-3\right)}{4\sin^{2}u},
\end{array}\vspace{3mm}\\
&&\label{ush-2-trigonometrico}
\begin{array}{l}
\mathcal{V}_2(u)=
-4\gamma^{2}{\cos^{4}u}+ 4\gamma\left(\gamma+2-2\beta \right){\cos^{2}u}+{4\left( \delta-\frac{1}{4}\right)
\left(\delta-\frac{3}{4}\right)\cot^2u}+\hspace{.9cm}\end{array}\nonumber\\
&&\hspace{1,8cm}\begin{array}{l}
{4\left( \beta+\delta+\ell-\frac{1}{4}\right)\left(\beta+\delta+\ell-\frac{3}{4}\right)\tan^2u},
\end{array}
%
%
%
\end{eqnarray}
where $\beta$,  $\gamma$,  $\delta$ and $\ell$  are real parameters.  According to Ushveridze, 
these are QES potentials because when $\ell$ is a natural number it is possible to find a finite number of energy levels 
and the respective wavefunctions. As  the
Schr\"{o}dinger equation reduces to a CHE  for these cases
\cite{lea},  
the problem  will be approached by using solutions of the CHE. 
%

On the other hand, finite series obtained from solutions of the CHE  have already been 
established for the following  QES hyperbolic potentials, given by  Xie \cite{xie} and Downing 
\cite{downing}, respectively:
\begin{eqnarray}
%
%
%
&&\label{hiperbolico-II}
{\cal V}_3(u) = -\frac{v_1}{\cosh^6u}-\frac{v_2}{\cosh^4u} -\frac{v_3}{\cosh^2u},
\vspace{3mm}\\
&&\label{hiperbolico-I}
{\cal V}_4(u) = -a\; \frac{\mbox{sinh}^4u}{\mbox{cosh}^6u},
\end{eqnarray}
%
%
%
%
%
where $ a$, $v_1$, $v_2$  and $v_3$ are real parameters, being $a>0$ and $v_1>0$.  
If $v_1=0$, the  Schr\"{o}dinger 
equation becomes  a 
reduced confluent Heun equation (RCHE)   and does not represent a QES problem. 
We will find that the procedure applied to the trigonometric potentials  leads to new solutions to 
the hyperbolic potentials (\ref{hiperbolico-II}) and (\ref{hiperbolico-I}).

For the CHE we use the form given by Leaver \cite{leaver} ($\omega\neq0$), namely,
\begin{eqnarray}
\label{che}
z(z-z_{0})\frac{d^{2}U}{dz^{2}}+(B_{1}+B_{2}z)
\frac{dU}{dz}+
\left[B_{3}-2\eta
\omega(z-z_{0})+\omega^{2}z(z-z_{0})\right]U=0, \quad[ CHE],
\end{eqnarray}
where $z_0$, $B_{i}$, $\eta$ and $\omega$ are constants. 
Eq. (\ref{che})  is responsible as well  for the angular 
and radial parts of    
two-center problem of  quantum mechanics and sometimes is called 
generalized spheroidal wave equation \cite{leaver,wilson,liu}.  
The RCHE \cite{lea,lay,kazakov} is written as ($q\neq0$)
%
%
\begin{eqnarray}\label{rche}
\label{lindemann}
z(z-z_{0})\frac{d^{2}Y}{dz^{2}}+(B_{1}+B_{2}z)
\frac{dY}{dz}+
\left[B_{3}+q(z-z_{0})\right]Y=0,\qquad [RCHE],
\end{eqnarray}
where $z_0$, $B_{i}$  and $q$ are constants. This RCHE describes,  for example, 
the angular dependence of the Schr\"{o}dinger 
equation for an electron in the field of a point dipole  \cite{lea-2,levy}.

In  Eqs. (\ref{che}) and (\ref{rche}),  	
the points	 $z=0$ and $z=z_{0}$ are regular singularities with exponents ($0,1+B_{1}/z_{0}$)
and ($0,1-B_{2}-B_{1}/z_{0}$), respectively. The point  $z=\infty$ is an irregular singularity  where 
the behaviour of the solutions for the CHE and  RCHE follows from the
normal and subnormal Thom\'e solutions, respectively,
namely \cite{lea,olver},
%
%
%
\begin{eqnarray}\label{thome1}
\mbox{when } z\rightarrow \infty \begin{cases}
U(z)\sim e^{\pm i\omega z}\;z^{\mp i\eta-\frac{B_{2}}{2}}, \qquad [\mbox{CHE}],\vspace{2mm}\\
Y(z)\sim
e^{\pm 2i\sqrt{qz}}\;z^{\frac{1}{4}-\frac{B_{2}}{2}},\qquad [\mbox{RCHE}].
\end{cases}
\end{eqnarray}
%

%


The RCHE (\ref{rche}) is obtained from the CHE (\ref{che}) by a procedure 
called Whittaker-Ince  limit (or Ince limit, simply) given by \cite{lea,eu2}
\begin{eqnarray}\label{limits}
\omega\rightarrow 0\; \mbox{ and }\;
\eta\rightarrow
\infty, \ \mbox{such that }\  \ 2\eta \omega =-q.
\end{eqnarray}
These relations are useful to find solutions for the RCHE. Further,
 if $z_0=0$, Eqs. (\ref{che}) and (\ref{rche}) become double-confluent Heun equations (DCHEs)
 which are useful for studying intermolecular forces and  
scattering of ions by polarizable targets \cite{eu2}, for example.  In some cases  \cite{lea}, 
solutions for DCHEs come from the Leaver limit $z_0\rightarrow 0$ applied to solutions of 
Eqs. (\ref{che}) and (\ref{rche}). 

Alternative forms,  used in Refs. \cite{xie,downing}, are obtained by substitution of variables, 
\begin{eqnarray}\begin{array}{lll}
\frac{d^{2}W}{dz^{2}}+
\left[2i\omega-\frac{\frac{B_1}{z_0}}{z}+\frac{B_{2}+\frac{B_1}{z_0}}{z-z_0} \right]\frac{dW}{dz}+\left[\frac{
\mu}{z}+\frac{
\nu}{z-z_{0}}\right]W=0,\quad &U(z)=e^{i\omega z}W(z),\quad &[\mbox{CHE}], \vspace{2mm}\\
\frac{d^{2}Z}{dz^{2}}+
\left[-\frac{\frac{B_1}{z_0}}{z}+\frac{B_{2}+\frac{B_1}{z_0}}{z-z_0} \right]\frac{dZ}{dz}+\left[  \frac{
q-\frac{B_3}{z_0}}{z}+\frac{
\frac{B_3}{z_0}}{z-z_{0}}  \right]Z=0,\quad &Y(z)=e^{i\omega z}Z(z),&[\mbox{RCHE}],
\end{array}
\end{eqnarray}
where $\mu=-\left(B_3+i\omega B_1+2\eta\omega z_0\right)/{z_0}$
and $\nu= i\omega B_2+\left(B_3+i\omega B_1\right)/{z_0}$.
%


The CHE for the previous potentials,  like the angular equation
of the two-center problem,  does not include 
the irregular singularity in the range of variable $z$ because 
 $ z=\cos^2u$,  $ z=\tanh^2u$ or $z=\operatorname{sech}^2u$, as we will see.  
 By this reason, solutions will be constructed from a power 
series solution for the CHE, applied by Baber and Hass\'e \cite{baber} to the angular part
of the wavefuntions for the hydrogen molecular ion. 

%
%
The series 
coefficients  satisfy three-term recurrence relations which  are equivalent to a 
homogeneous system  represented by a tridiagonal matrix.
By the way, a QES problem is  the one  which
has solutions given 
by finite series whose coefficients satisfy  three-term  or higher order
recurrence relations \cite{Kalnins}. In the present case, 
the energy spectra  corresponding to finite series may be determined from the vanishing
of the matrix determinant (characteristic equation).
 For each energy we need to find the series coefficients.

Nevertheless, we will find that the  power expansions  allow as well infinite-series solutions
with a transcendental equation to determine the energies and the series coefficients. So,
despite the finite series, it is necessary to consider 
infinite series    
as in the case of the two-center problem.

%
To find the eigenfuctions, firstly we  present some facts 
concerning the power series expansions of  the CHE, their
convergence  and linear dependence. 
Thus, in Section 2, we see  that d'Alembert's ratio test implies 
that the series converge  for any finite 
value of $z$ and, in particular,
for $0\leq |z|\leq |z_0|$, as stated in Ref. \cite{baber}.  Such convergence is valid also for solutions of the  RCHE. 
This assures that  the infinite-series solutions converge for all values of $z$ 
in the above problems.

On the other side, by transformations of 
variables we generate
a group of 16 solutions. 
Thence, for a given finite-series solution,  we will  find a suitable infinite-series solution for the same set of parameters.  This clarifies the coexistence of finite and infinite series. 

Sections 3 and 4 supply finite- and infinite-series wavefunctions for 
Ushveridze's trigonometric potentials (\ref{ush-1-trigonometrico}) and (\ref{ush-2-trigonometrico}), 
respectively. 
In general  it is necessary to assign values for the parameters of the potentials. The particular 
cases considered in Secs, 3.1 and 4.1 revel the efficiency and some details of the procedure. Further,
at the end of such sections we indicate other trigonometric potentials that can be solved in the same way.

Since the hyperbolic potentials  (\ref{hiperbolico-II})  and (\ref{hiperbolico-I}) have already 
been treated in the context of the CHE,
in Sections 5 and 6 we discuss only the possibility of new finite-series solutions
resulting  from the 16 expansions for the CHE, as well as
the existence of infinite-series  solutions. Sec. 6.1 
considers the RCHE corresponding
to $v_1=0$ in Eq. (\ref{hiperbolico-II}). Some concluding remarks are in Sec. 7.

%
%
%
\section{Power series solutions for the CHE}
\indent
In this section we write the four basic 
transformations $T_i$ for  the CHE and use them to generate a group of 16
solutions in series of $z$ and $z-z_0$ having series coefficients $b_n$ which satisfy 
three-term recurrence relations. We discuss also the convergence of infinite series and the 
linear dependence.

\subsection{Transformations of the CHE}

If we set $U(z)=U(B_{1},B_{2},B_{3}; z_{0},\omega,\eta;z)$
%
%
%
in the CHE (\ref{che}), the form of the equation is preserved  by four  
transformations of variables  and parameters \cite{lea,decarreau2}, namely, 
%
\begin{eqnarray}\label{trans}
\begin{array}{l}
T_{1}U(z)=z^{1+\frac{B_{1}}{z_{0}}}
U(C_{1},C_{2},C_{3};z_{0},\omega,\eta;z),\  \ z_{0}\neq0,
\vspace{3mm}\\
T_{2}U(z)=(z-z_{0})^{1-B_{2}-\frac{B_{1}}{z_{0}}}U(B_{1},D_{2},D_{3};
z_{0},\omega,\eta;z), \ \  z_{0}\neq0,
\vspace{3mm}\\
T_{3}U(z)=U(B_{1},B_{2},B_{3}; z_{0},-\omega,-\eta;z),
\ \ \forall z_{0},
\vspace{3mm}\\
T_{4}U(z)=
U(-B_{1}-B_{2}z_{0},B_{2},
B_{3}+2\eta\omega z_{0};z_{0},-\omega,
\eta;z_{0}-z),\ \  \forall z_{0}
\end{array}
\end{eqnarray}
where
\begin{eqnarray}
\label{constantes-C-D}
\begin{array}{l}
C_{1}=-B_{1}-2z_{0}, \ \
C_{2}=2+B_{2}+\frac{2B_{1}}{z_{0}},\ C_{3}=B_{3}+
\left(1+\frac{B_{1}}{z_{0}}\right)
\left(B_{2}+\frac{B_{1}}{z_{0}}\right),
\vspace{.3cm}\\
D_{2}=2-B_{2}-\frac{2B_{1}}{z_{0}},\ D_{3}=B_{3}+
\frac{B_{1}}{z_{0}}\left(\frac{B_{1}}{z_{0}}
+B_{2}-1\right).
\end{array}
\end{eqnarray}
%
%

On the other side,  the  form of the recurrence relations  for  the coefficient 
$b_{n}$ is \cite{leaver}
\begin{eqnarray}\label{recurrence1}
\alpha_{0}b_{1}+\beta_{0}b_{0}=0,
\qquad \alpha_{n}b_{n+1}+\beta_{n}b_{n}+
\gamma_{n}b_{n-1}=0\quad (n\geq1),
\end{eqnarray}
where $\alpha_{n}$, $\beta_{n}$ and $\gamma_{n}$ 
depend on the parameters of the
differential equation. Relations (\ref{recurrence1}) represent 
an infinite system of homogeneous linear equations associated with a tridiagonal matrix 
whose deteminant must vanish (characteristic equation) 
in order to get non-trivial solutions.  This is possible only if there is a free parameter
in the CHE, as ${\cal E}$ in Eq. (\ref{schr}). The  solutions are
written as infinite series ($n$ running from $0$ to $\infty$), but  these series 
can terminate on the right-hand side provided that $\gamma_n=0$ for some 
natural number $n=N+1\geq 1$ 
\cite{arscott} and, then, we have
\begin{eqnarray}\label{truncation}
\begin{array}{l}
 \text{finite series with } 0\leq n\leq N \text{ if }\gamma_{N+1}=0.
\end{array}
\end{eqnarray}

By applying the previous transformations to an inicial solution $U^{(1)}(z)$, we 
generate a group containing 16 power-series  expansions for 
the CHE: eight expansions denoted by $U^{(i)}(z)$ are given by series of 
$z$, while eight $\bm{U}^{(i)}(z)$ are given by series of 
$z-z_0$. According to (\ref{U1}) and (\ref{UU1}),  the forms of $U^{(1)}(z)$ and $\bm{U}^{(1)}(z)$ are
\begin{eqnarray}
U^{(1)}(z)=e^{i\omega z}\displaystyle \sum_{n=0}^{\infty}b_{n}^{(1)}
z^{n},\qquad 
\bm{U}^{(1)}(z)=T_4U^{(1)}(z)=e^{i\omega z}\displaystyle 
\sum_{n=0}^{\infty}\bm{b}_{n}^{(1)}(z-z_0)^n,
\end{eqnarray}
where $U^{(1)}(z)$ is a Baber and Hass\'e solution \cite{baber} discussed by Leaver \cite{leaver}.
From these, the other expansions follow by using  $T_{1}$, $T_{2}$ and $T_3$ as 
\begin{align}
&\label{t}
 \begin{array}{l}
U^{(1)},\quad
U^{(2)}=T_1U^{(1)} ,\quad\;
U^{(3)}=T_2U^{(2)},\quad\;
U^{(4)}=T_1U^{(3)},\quad U^{(i+4)}=T_3U^{(i)};
\end{array}\vspace{2mm}\\
&\label{TT}
\begin{array}{l}
\bm{U}^{(1)},\quad
\bm{U}^{(2)}=T_1\bm{U}^{(1)} ,\quad
\bm{U}^{(3)}=T_2\bm{U}^{(2)},\quad
\bm{U}^{(4)}=T_1\bm{U}^{(3)},\quad \bm{U}^{(i+4)}=T_3\bm{U}^{(i)};
%
\end{array}
\end{align}
where $i =1,2,3,4$. 
Since the transformation $T_3$ simply replace $(\omega,\eta)$ with  
$(-\omega,-\eta)$, we do not write explicitly the expansions ${U}^{(i+4)}$ and  $\bm{U}^{(i+4)}$.

%

Further, if $U(z)=U(B_{1},B_{2},B_{3};z_{0},q;z)$ 
represents an arbitrary solution for the RCHE  (\ref{rche}), then other
solutions are generated by the transformations
$\mathscr{T}_1$, $\mathscr{T}_2$ and $\mathscr{T}_3$
given by \cite{lea}
\begin{eqnarray}\label{Transformacao2}
\begin{array}{l}
\mathscr{T}_{1}
U(z)=z^{1+B_{1}/z_{0}}
U(C_{1},C_{2},C_{3};z_{0},q;z),\vspace{2mm}\\
\mathscr{T}_{2}
U(z)=(z-z_{0})^{1-B_{2}-B_{1}/z_{0}}U(B_{1},D_{2},D_{3};
z_{0},q;z), \vspace{2mm}\\
\mathscr{T}_{3}
U(z)=
U(-B_{1}-B_{2}z_{0},B_{2},
B_{3}-q z_{0};z_{0},-q;z_{0}-z),
\end{array}
\end{eqnarray}
where $C_{i}$ and $D_{i}$ are defined in Eqs. (\ref{constantes-C-D} ).
These  give at most 8 expansions for the RCHE.

\subsection{Series of $z$}
The Baber and Hass\'e expansion $U^{(1)}(z)$ is used as the initial solution. 
For Leaver's form (\ref{che}) of the
CHE, it  reads  \cite{leaver}
\begin{eqnarray}\label{U1}
\begin{cases}
U^{(1)}(z)=e^{i\omega z}\displaystyle \sum_{n=0}^{\infty}b_{n}^{(1)}
z^{n},\\
\alpha_n^{(1)}=- z_{0}\left(n-\frac{B_{1}}{z_{0}}\right)
\left(n+1\right), \qquad \beta_n^{(1)}=
n\left(n+B_{2}-1-2i\omega z_{0}\right)+\\
%
B_{3}+2\eta\omega z_0+ i\omega B_{1},\qquad \gamma_n^{(1)}=
2i\omega\left(n+i\eta+\frac{B_{2}}{2}-1\right).
\end{cases}
\end{eqnarray}
The factor $\exp{(i\omega z)}$ assures that the coefficients $b_n^{(1)}$ satisfy 
the three-term recurrence relations (\ref{recurrence1}).
The transformations  (\ref{t}) lead to the  other expansions
$U^{(i)}(z)$.
\begin{eqnarray}
\begin{cases}
U^{(2)}(z)=e^{i\omega z}z^{1+\frac{B_1}{z_0}}\displaystyle \sum_{n=0}^{\infty}b_{n}^{(2)}
z^{n},\vspace{2mm}\\
\alpha_n^{(2)}=- z_{0}\left(n+2+\frac{B_{1}}{z_{0}}\right)
\left(n+1\right), \qquad
\beta_n^{(2)}=
n\left(n+1+B_{2}+\frac{2B_1}{z_0}-2i\omega z_{0}\right)+\vspace{2mm}\\
B_{3}+2\eta\omega z_0-i\omega(B_1+2z_0)+
\left(B_2+\frac{ B_1}{z_0}\right)
\left(1+\frac{B_1}{z_0}\right),
\qquad 
\gamma_n^{(2)}=
2i\omega\left(n+i\eta+\frac{B_1}{z_0}+\frac{B_{2}}{2}\right).
\end{cases}
\end{eqnarray}
\begin{eqnarray}
\begin{cases}
U^{(3)}(z)=e^{i\omega z}z^{1+\frac{B_{1}}{z_{0}}}
(z-z_{0})^{1-B_{2}-\frac{B_{1}}{z_{0}}}\displaystyle \sum_{n=0}^{\infty}b_{n}^{(3)}
z^{n},\\
\alpha_n^{(3)}=- z_{0}\left(n+2+\frac{B_{1}}{z_{0}}\right)
\left(n+1\right), \qquad \beta_n^{(3)}=
n\left(n+3-B_{2}-2i\omega z_{0}\right)+\\
2-B_2+B_{3}+2\eta\omega z_0- i\omega (B_{1}+2z_0),\qquad \gamma_n^{(3)}=
2i\omega\left(n+i\eta+1-\frac{B_{2}}{2}\right).
\end{cases}
\end{eqnarray}
\begin{eqnarray}
\begin{cases}
U^{(4)}(z)=e^{i\omega z}(z-z_{0})^{1-B_{2}-\frac{B_{1}}{z_{0}}}
\displaystyle \sum_{n=0}^{\infty}b_{n}^{(4)}
z^{n},\\
\alpha_n^{(4)}=- z_{0}\left(n-\frac{B_{1}}{z_{0}}\right)
\left(n+1\right), \qquad \beta_n^{(4)}=
n\left(n+1-B_{2}-\frac{2B_1}{z_0}-2i\omega z_{0}\right)+\\
%
B_{3}+2\eta\omega z_0+ i\omega B_{1}+\frac{B_1}{z_0}\left(\frac{B_1}{z_0}+B_2-1\right),\qquad \gamma_n^{(4)}=
2i\omega\left(n+i\eta-\frac{B_1}{z_0}-\frac{B_{2}}{2}\right).
\end{cases}
\end{eqnarray}

\subsection{Series of $z-z_0$}

We find that in the recurrence relations for ${U}^{(i)}(x)$ and  $\bm{U}^{(i)}(x)$
\begin{eqnarray*}
\gamma_n^{(i)}=\bm\gamma_n^{(i)}, \qquad[i=1,2,\cdots, 8]
\end{eqnarray*}
Then, if ${U}^{(i)}(x)$ is an expansion in finite or infinite series, the same is true for $\bm{U}^{(i)}(x)$
due to (\ref{truncation}).
\begin{eqnarray}\label{UU1}
\begin{cases}
\bm{U}^{(1)}(z)=e^{i\omega z}\displaystyle \sum_{n=0}^{\infty}\bm{b}_{n}^{(1)}
(z-z_{0})^{n},\vspace{2mm}\\
\bm\alpha_n^{(1)}= z_{0}\left(n+B_{2}+\frac{B_{1}}{z_{0}}\right)
\left(n+1\right), \qquad \bm\beta_n^{(1)}=
n\left(n+B_{2}-1+2i\omega z_{0}\right)+\vspace{2mm}\\
B_{3}+ i\omega z_{0}\left(B_{2}+
\frac{B_{1}}{z_{0}}\right), \qquad 
\bm\gamma_n^{(1)}=
2i\omega\left(n+i\eta+\frac{B_{2}}{2}-1\right)=\gamma_n^{(1)}.
\end{cases}
\end{eqnarray}
\begin{eqnarray}
\begin{cases}
\bm{U}^{(2)}(z)=e^{i\omega z}z^{1+\frac{B_{1}}{z_{0}}}
\displaystyle \sum_{n=0}^{\infty}\bm{b}_{n}^{(2)}(z-z_{0})^{n},
\vspace{2mm}\\
%
%
\bm\alpha_n^{(2)}=z_{0}\left(n+B_{2}+\frac{B_{1}}{z_{0}}\right)(n+1),
\qquad \bm\beta_n^{(2)}=n\left[n+1+2i\omega z_{0}+B_{2}
+\frac{2B_{1}}{z_{0}}\right]+B_3
\vspace{2mm}\\
+i\omega z_{0}\left[B_{2}+\frac{B_{1}}{z_{0}}\right]
+
\left[1+\frac{B_{1}}{z_{0}}\right]\left[B_{2}+
\frac{B_{1}}{z_{0}}\right],\qquad \bm\gamma_n^{(2)}=
2i\omega \left[n+i\eta+\frac{B_{1}}{z_{0}}+\frac{B_{2}}{2}\right]=\gamma_n^{(2)}.
\end{cases}
\end{eqnarray}
%
%
\begin{eqnarray}\label{barber-7}
\begin{cases}\bm{U}^{(3)}(z)
=e^{i\omega z}z^{1+\frac{B_{1}}{z_{0}}}
(z-z_{0})^{1-B_{2}-\frac{B_{1}}{z_{0}}}
\displaystyle \sum_{n=0}^{\infty}\bm{b}_{n}^{(3)}(z-z_{0})^{n},\vspace{2mm}\\
%
%
\bm\alpha_n^{(3)}=z_{0}\left(n+2-B_{2}-\frac{B_{1}}{z_{0}}
\right)(n+1), \qquad 
\bm\beta_n^{(3)}=
n\left(n+3+2i\omega z_{0}-B_{2}\right)
+2-B_{2}+
\vspace{2mm}\\
%
%
B_{3}+i\omega z_{0} \left(2-B_{2}-\frac{B_{1}}{z_{0}}\right),
\qquad
\bm\gamma_n^{(3)}=
2i\omega \left(n+1+i\eta-\frac{B_{2}}{2}\right)=\gamma_n^{(3)} .
\end{cases}
\end{eqnarray}
\begin{eqnarray}\label{baber-8}
\begin{cases}\bm{U}^{(4)}(z)
=e^{i\omega
 z}(z-z_{0})^{1-B_{2}-\frac{B_{1}}{z_{0}}}
\displaystyle \sum_{n=0}^{\infty}\bm{b}_{n}^{(4)}(z-z_{0})^{n},\vspace{2mm}\\
%
%
\bm\alpha_n^{(4)}=z_{0}\left(n+2-B_{2}-\frac{B_{1}}{z_{0}}\right)(n+1),\qquad
\bm\beta_n^{(4)}=
n\left(n+1+2i\omega z_{0}-B_{2}-
\frac{2B_{1}}{z_{0}}\right)+B_3+\vspace{2mm}\\
%
%
i\omega z_{0}
\left(2-B_{2}-\frac{B_{1}}{z_{0}}\right)+
\frac{B_{1}}{z_{0}}\left(B_{2}+\frac{B_{1}}{z_{0}}-1\right)
,\qquad
\bm\gamma_n^{(4)}=2i\omega \left(n+i\eta-\frac{B_{1}}{z_{0}}-\frac{B_{2}}{2}\right)=\gamma_n^{(4)}.
\end{cases}
\end{eqnarray}

\subsection{Convergence of infinite series}

Relations (\ref{recurrence1}) in general have two 
independent solutions \cite{gautschi}  but 
only one of theses assure the convergence 
of the characteristic equation. 
For  example, the recurrence relations of the expansion $U^{(1)}(z)$ divided by
$n^ 2\;{b}_ n^{(1)}$ give
%
%
%
%
%
 %
%
%
 %
\begin{eqnarray*}
&\begin{array}{l}
z_0\left[1+\frac{1}{n}\left(1-\frac{B_1}{z_0}\right)-\frac{1}{n^2}\frac{B_1}{z_0}\right]
\frac{{b}_{n+1}^{(1)}}{{b}_n^{(1)}}-
\left[1+\frac{1}{n}\left(B_{2}-1-2i\omega z_{0}\right)+\frac{1}{n^2}(B_{3}+2\eta\omega z_0+ i\omega B_{1})\right]
\end{array}\vspace{2mm}\\
&\begin{array}{l}
-2i\omega\left[\frac{1}{n}+\frac{1}{n^2}\left(i\eta+\frac{B_{2}}{2}-1\right)\right]
\frac{{b}_{n-1}^{(1)}}{{b}_n^{(1)}}=0.
\end{array}
\end{eqnarray*}
Then,  for large $n$  the two ratios  for 
${b}_{n+1}^{(1)}/{b}_n^{(1)}$ are 
\begin{eqnarray} 
&&\begin{array}{l}\label{minimal-0}
  \frac{{b}_{n+1}^{(1)}}{{b}_n^{(1)}}\sim-\frac{2i\omega}{n}\left[
  1+\frac{1}{n}\left(i\eta-\frac{B_2}{2}\right)
  \right]\Rightarrow 
  \frac{{b}_{n-1}^{(1)}}{{b}_n^{(1)}}\sim-\frac{n}{2i\omega}\left[
  1-\frac{1}{n}\left(i\eta-\frac{B_2}{2}\right)
  \right],\end{array}\vspace{2mm}\\
&&\begin{array}{l}\label{dominant-0}
  \frac{{b}_{n+1}^{(1)}}{{b}_n^{(1)}}\sim 
\frac{1}{z_0}\left[1+\frac{1}{n}\left(B_2+\frac{B_1}{z_0}-1\right)\right] \Rightarrow
\frac{{b}_{n-1}^{(1)}}{{b}_n^{(1)}}\sim
z_0\left[1-\frac{1}{n}\left(B_2+\frac{B_1}{z_0}-1\right)\right] ,
\end{array}
\end{eqnarray}
which imply respectively
\begin{eqnarray} 
&&\begin{array}{l}\label{minimal}
\vline\frac{{b}_{n+1}^{(1)}z^{n+1}}{{b}_{n}^{(1)}z^{n}}\vline\sim\;
\vline\frac{2i\omega z}{n}\left[1+\frac{1}{n}\left(i\eta-\frac{B_2}{2}\right)\right]\vline
  \end{array}\vspace{2mm}\\
&&\begin{array}{l}\label{dominant}
 \vline\frac{{b}_{n+1}^{(1)}z^{n+1}}{{b}_{n}^{(1)}z^{n}}\vline\sim\;
\vline\frac{z}{z_0}\left[1+\frac{1}{n}\left(B_2+\frac{B_1}{z_0}-1\right)\right]\vline .
\end{array}
\end{eqnarray}
However, (\ref{minimal-0}) and  (\ref{dominant-0}) give $\displaystyle\lim_{n\to\infty}{b}_{n+1}^{(1)}\big/{b}_n^{(1)}=0$ and 
$\displaystyle\lim_{n\to\infty}{b}_{n+1}^{(1)}\big/{b}_n^{(1)}=1/z_0$, respectively.  Since these limits are distinct, by a Poincar\'e-Perron 
theorem we have to choose the first case in order to assure the convergence  of the 
characteristic equation, given by the determinant of an infinite matrix or by
an infinite continued fraction \cite{gautschi}. Therefore,
%
%
%
%
%
%
%
\begin{eqnarray}
\label{conve-1}
\vline\frac{{b}_{n+1}^{(1)}z^{n+1}}{{b}_{n}^{(1)}z^{n}}\vline\sim\;
\vline
\frac{2i\omega z}{n}\vline \quad \mbox{ for the CHE};\qquad
\vline\frac{{b}_{n+1}^{(1)}z^{n+1}}{{b}_{n}^{(1)}z^{n}}\vline\sim\;
\vline\frac{q z}{n^2}\vline\qquad \mbox{ for the RCHE}.
\end{eqnarray}
Consequently, $U^{(1)}(z)$ converges for any finite values of $z$ provided that $n$ is
suficiently large.  In particular, the solution converges in the region $0\leq |z|\leq |z_0|$
as stated after Eq. (10) of the paper by Baber and Hass\'e \cite{baber}. Transformations (\ref{trans})
do not modify this fact.


%
%
%

%
\subsection{Linear dependence and finite series}

If $U^{(i)}(z)$ and 
$\bm{U}^{(i)}(z)$ are given by infinite series ($\gamma_{\;n}^{(i)}=\bm\gamma_{\;n}^{(i)}\neq0$),
 we have the relations 
%
%
\begin{align}
&\label{a1}\begin{array}{l}
\bm{U}^{(i)}(z)=
{U}^{(i)}(z),
	\quad i=1,2,\cdots,8;
 \end{array}
\vspace{3mm}\\
&\label{a2}\begin{array}{r}
U^{(1)}(z)=U^{(4)}(z),\quad U^{(2)}(z)=U^{(3)}(z),\quad
U^{(5)}(z)=U^{(8)}(z),\quad U^{(6)}(z)=U^{(7)}(z),
%
\end{array}
\vspace{3mm}\\
%
%
%
&\label{a3}\begin{array}{l}
{U}^{(i+4)}(z)=
	{U}^{(i)}(z),
	\;(i=1,2,3,4)\Rightarrow {U}^{(5)}=
	{U}^{(1)}, \; {U}^{(6)}=
	{U}^{(2)}, \; {U}^{(7)}=
	{U}^{(3)}, \; {U}^{(8)}=
	{U}^{(4)}.
\end{array}
\end{align}
In Section 3.1, there are cases in which  $U^{(i)}(z)$ and 
$\bm{U}^{(i)}(z)$ are all given by infinite series; the above relations will imply that only
two expansions are linearly independent since
\begin{eqnarray}\label{a4}
	{U}^{(1)}(z)= {U}^{(4)}(z)=
	{U}^{(5)}(z)= {U}^{(8)}(z),
	\qquad{U}^{(2)}(z)= {U}^{(3)}(z)=
	{U}^{(6)}(z)= {U}^{(7)}(z),
\end{eqnarray}


%
%
%
%
%
%
%
%
%
Relation (\ref{a1}) is obtained by taking $x=z$ and $a=-z_0$ in the  the binomial series \cite{Gradshteyn}
%
%
\begin{eqnarray*}
(x+a)^n=
\sum_{l=0}^{n}\frac{n!\;x^n\;a^{n-l}}{l!\;(n-l)!}.
\end{eqnarray*}
%
This implies
\begin{eqnarray}\label{binomio}
\sum_{n=0}^\infty \bm{b}_n^{(i)}(z-z_0)^n=
\sum_{n=0}^\infty z^n\left[\bm{b}_n^{(i)}\;
n!\sum_{l=0}^{n}\frac{(-z_0)^{n-l}}{l!\;(n-l)!}\right]=
\sum_{n=0}^{\infty}b_n^{(i)} z^n, 
\end{eqnarray}
which give the connection between the coefficients $\bm{b}_n^{(i)}$ and
${b}_n^{(i)}$  of expansions  $\bm{U}^{(i)}$ and
${U}^{(i)}$.
%
%

On the other hand, in each equality given in (\ref{a2}), the expansions differ by the factor $(z-z_{0})^{1-B_{2}-\frac{B_{1}}{z_0}}$
which can be rewritten as 
%
%
\begin{eqnarray}\label{aqui}
\begin{array}{l}
(z-z_{0})^{1-B_{2}-\frac{B_{1}}{z_0}}
=C\displaystyle\sum_{m=0}^\infty 
\frac{(-\frac{z}{z_0})^m}
{m!\;\Gamma\left(2-B_2-\frac{B_1}{z_0}-m\right)},
\qquad B_2+\frac{B_1}{z_0}\neq 
0, -1,-2,\cdots,
\end{array}
\end{eqnarray}
where $C=(-z_{0})^{1-B_{2}-\frac{B_{1}}{z_{0}}}\Gamma\left(2-B_2-\frac{B_1}{z_0}\right)$. 
The above relation results from the fact that, 
if $q$ is not a natural number,  we have the infinite series \cite{Gradshteyn}
\begin{eqnarray*}\label{binomial}
(1+y)^{q}&=&\sum_{m=0}^\infty \frac{\Gamma(q+1)}{m!\;\Gamma(q+1-m)}\;y^m\nonumber\\
&=&1+qy+ \frac{q(q-1)}{2!}y^2+\cdots+\frac{q(q-1)\cdots(q-k+1)}{k!}y^k+\cdots ,
\end{eqnarray*}
which converges absolutely for $|y|<1$ and diverges for $|y|>1$. Further  \cite{Gradshteyn}
\begin{eqnarray}\label{cauchy}
\sum_{n=0}^\infty a_n x^n\sum_{n=0}^\infty b_n x^n = \sum_{n=0}^\infty c_n x^n,
\quad\mathrm{where}\qquad c_n=\sum_{l=0}^n a_l\, b_{n-l}.
\end{eqnarray}
%
%
Hence, from (\ref{aqui}) and (\ref{cauchy}), we obtain 
\begin{eqnarray}
\begin{array}{l}
(z-z_{0})^{1-B_{2}-\frac{B_{1}}{z_{0}}}
\displaystyle\sum_{n=0}^\infty b_n z^n
=C\displaystyle\sum_{n=0}^\infty z^n
\left[
\displaystyle\sum_{l=0}^n 
\frac{(-z_0)^{-l}\;b_{n-l}}
{l!\;\Gamma\left(2-B_2-\frac{B_1}{z_0}-l\right)}
\right],
\end{array}
\end{eqnarray}
which can be used to prove (\ref{a2}) by redefinitions of series coefficients. Finally, we can
write
%
%
%
\begin{eqnarray}
e^{-x}=e^{x}e^{-2x}=e^{x}\sum_{n=0}^\infty\frac{(-2)^n}{n!}x^n.
\end{eqnarray}
and use this together with  relation (\ref{cauchy}) to establish (\ref{a3}).


In fact, for infinite series the parameters of the CHE are such that
$\gamma_{n}\neq 0$ in the recurrence relations $\alpha_{n}b_{n+1}+\beta_{n}b_{n}+
\gamma_{n}b_{n-1}=0$ ($b_{-1}=0$). On the other side,  if 
$\gamma_{N+1}=0$, we obtain a finite series with $N+1$ terms ($0\leq n\leq N$), as stated in Eq. (\ref{truncation}). 
Then, if $\beta_i=\mathcal{B}_i-\Lambda$ ($i=0,1,\cdots,N$) 
and if $\alpha_i$, $\mathcal{B}_i$
and $\gamma_i$ are real and independent of the constant $\Lambda$, then \cite{arscott}
%
%
%
%
\begin{eqnarray}\label{autovalores}
\mbox{finite series ($0\leq n\leq N$ $)\Rightarrow $ $N+1$ real 
and distinct values for }\Lambda \;
\mbox{ if }\;\alpha_{n-1}\;\gamma_n>0, 
\\
\left[  \text{Arscott conditions}\right].
\nonumber
\end{eqnarray} 
%
For finite series concerning 
the trigonometric potentials (\ref{ush-1-trigonometrico}) and (\ref{ush-2-trigonometrico}), 
$U^{(i)}$ and $\bm{U}^{(i)}$ will give solutions which 
satisfy conditions (\ref{autovalores}) for opposite signs of $\beta$ or $\gamma$.
However, the finite series for 
the hyperbolic potentials (\ref{hiperbolico-I}) and (\ref{hiperbolico-II}) will not satisfy conditions required by the theorem.

\subsection{Limits for the  reduced CHE}

The  limits (\ref{limits})  give only the following four infinite-series expansions $Y^{(i)}$ for the RCHE. The transformation $\mathscr{T}_{3}$ written in (\ref{Transformacao2}) does not produce 
new expansions in series of $z-z_0$ due to (\ref{binomio}).
%
%
%
\begin{eqnarray}
\begin{cases}
Y^{(1)}(z)=\displaystyle \sum_{n=0}^{\infty}b_{n}^{(1)}
z^{n},\\
\alpha_n^{(1)}=- z_{0}\left(n-\frac{B_{1}}{z_{0}}\right)
\left(n+1\right), \quad \beta_n^{(1)}=
n\left(n+B_{2}-1\right)+
%
B_{3}-q z_0,\qquad \gamma_n^{(1)}=q.
\end{cases}
\end{eqnarray}
%
%
\begin{eqnarray}
\begin{cases}
Y^{(2)}(z)=z^{1+\frac{B_{1}}{z_{0}}}\displaystyle \sum_{n=0}^{\infty}b_{n}^{(2)}
z^{n},\vspace{2mm}\\
\alpha_n^{(2)}=- z_{0}\left(n+2+\frac{B_{1}}{z_{0}}\right)
\left(n+1\right), \qquad 
\beta_n^{(2)}=
n\left(n+1+B_{2}+\frac{2B_1}{z_0}\right)+\vspace{2mm}\\
B_{3}-qz_0+
\left(B_2+\frac{ B_1}{z_0}\right)
\left(1+\frac{B_1}{z_0}\right),
\qquad 
\gamma_n^{(2)}=q.
\end{cases}
\end{eqnarray}
\begin{eqnarray}
\begin{cases}
Y^{(3)}(z)=z^{1+\frac{B_{1}}{z_{0}}}
(z-z_{0})^{1-B_{2}-\frac{B_{1}}{z_{0}}}\displaystyle \sum_{n=0}^{\infty}b_{n}^{(3)}
z^{n},\\
\alpha_n^{(3)}=- z_{0}\left(n+2+\frac{B_{1}}{z_{0}}\right)
\left(n+1\right), \quad \beta_n^{(3)}=
n\left(n+3-B_{2}\right)+
2-B_2+B_{3}-qz_0,\quad \gamma_n^{(3)}=q.
\end{cases}
\end{eqnarray}
\begin{eqnarray}
\begin{cases}
Y^{(4)}(z)=(z-z_{0})^{1-B_{2}-\frac{B_{1}}{z_{0}}}
\displaystyle \sum_{n=0}^{\infty}b_{n}^{(4)}
z^{n},\\
\alpha_n^{(4)}=- z_{0}\left(n-\frac{B_{1}}{z_{0}}\right)
\left(n+1\right), \quad \beta_n^{(4)}=
n\left(n+1-B_{2}-\frac{2B_1}{z_0}\right)+\\
%
B_{3}-q z_0+\frac{B_1}{z_0}\left(\frac{B_1}{z_0}+B_2-1\right),\quad \gamma_n^{(4)}=
q.
\end{cases}
\end{eqnarray}

\section{The trigonometric potential given in  Eq. (\ref{ush-1-trigonometrico})}

For the trigonometric potential (\ref{ush-1-trigonometrico}),
the Schr\"{o}dinger equation (\ref{schr}) becomes
  \begin{eqnarray}
  \label{schr-2a}
&&
\frac{d^2\psi}{du^2}+
\left\{ {\cal E}+
  4\beta^{2}{\sin^{4}u}- 4\beta\big[\beta+
 2(\gamma+\delta+\ell) \big]{\sin^{2}u}-
 {4\left( \gamma-\frac{1}{4}\right) \left(\gamma-\frac{3}{4}
    \right)}
 \operatorname{sec}^{2}u  \right.\nonumber\vspace{2mm}\\
&&
\left.
 -
 {4\left( \delta-\frac{1}{4}\right) \left(\delta-\frac{3}{4}\right)}
  \csc^2u
  \right\}
  \psi=0. 
\end{eqnarray}
%
 %
%
%
The substitutions
 \begin{eqnarray}\label{trig-1}
  z=\cos^2u,\qquad \psi(u)={\psi}[u(z)]=
  z^{\delta-\frac{1}{4}}(z-1)^{\gamma-\frac{1}{4}}U(z),
  \qquad\left[0\leq z\leq 1\right] 
   \end{eqnarray}
 transform the above equation into a confluent Heun equation for $U(z)$,
 \begin{eqnarray}
 \label{gswe-1}
&&\begin{array}{l}
z(z-1)\frac{d^{2}U}{dz^{2}}+[-2\gamma+(2\gamma+2\delta)z]
\frac{dU}{dz}+\Big[\left(\gamma+\delta-\frac{1}{2}\right)^{2}-\frac{{\cal E}}{4}
+2\beta(\gamma+\delta+\ell)(z-1)
\end{array}\nonumber\vspace{2mm}\\
&&\begin{array}{l}
-\beta^{2}z(z-1)\Big]U=0.
\end{array} 
\end{eqnarray}
%
Thence, the parameters of the CHE  (\ref{che}) are ($z_0=1$)
 \begin{eqnarray}\label{trig-a}
\begin{array}{l}
B_{1}=-2\delta, \quad  B_{2}=2\gamma+2\delta, \quad 
 B_{3}=\left(\gamma+\delta-\frac{1}{2}\right)^{2}-\frac{{\cal E}}{4},\quad
 %
i\omega=\beta,\quad i\eta= -(\ell+\gamma+\delta).
 \end{array}
 \end{eqnarray}

  The wavefunctions obtained by inserting  $\left({U}^{(i)},\bm{U}^{(i)}\right)$ 
  into  (\ref{trig-1})  
  are denoted by $\left({\psi}_{\;\ell}^{(i)},\bm{\psi}_{\;\ell}^{(i)}\right)$ and must be 
  bounded for any value of $z$  and, in particular, when $z=1$ and $z=0$. This requirement
  imposes restrictions on the values of $\gamma$ and $\delta$, namely,
  %
  %
\begin{eqnarray} \label{gamma-delta}
\begin{array}{l}
\left({\psi}_{\;\ell}^{(1)},\bm{\psi}_{\;\ell}^{(1)}\right) ,\left({\psi}_{\;\ell}^{(5)},\bm{\psi}_{\;\ell}^{(5)}\right): \gamma\geq\frac{1}{4}, \;\delta\geq\frac{1}{4};\quad
\left({\psi}_{\;\ell}^{(2)},\bm{\psi}_{\;\ell}^{(2)}\right) ,\left({\psi}_{\;\ell}^{(6)},\bm{\psi}_{\;\ell}^{(6)}\right): \gamma\geq\frac{1}{4}, \;\delta\leq\frac{3}{4};\vspace{3mm}\\
\left({\psi}_{\;\ell}^{(3)},\bm{\psi}_{\;\ell}^{(3)}\right),\left({\psi}_{\;\ell}^{(7)},\bm{\psi}_{\;\ell}^{(7)}\right): \gamma\leq\frac{3}{4}, \;\delta\leq\frac{3}{4};\quad
\left({\psi}_{\;\ell}^{(4)},\bm{\psi}_{\;\ell}^{(4)}\right),\left({\psi}_{\;\ell}^{(8)},\bm{\psi}_{\;\ell}^{(8)}\right): \gamma\leq\frac{3}{4}, \;\delta\geq\frac{1}{4}.
\end{array}
\end{eqnarray}
These restrictions assure that the exponents of  $z-1$ and $z$, respectively, are not negative numbers.   Besides this, the presence of two pairs of solutions  for each range of $\gamma$ and 
 $\delta$ is a fact relevant to get finite and infinite-series solutions at the same time.
 The general outcome is:
%
   %
   \begin{eqnarray}\label{prescription}
   \begin{array}{ll}
\mbox{if  }\left({U}^{(i)},\bm{U}^{(i)}\right) 
 \mbox{ gives finite series, }&\mbox{then } \left({U}^{(i+4)},\bm{U}^{(i+4)}\right) 
 \mbox{ gives infinite series};\vspace{2mm}\\
 \mbox{if } \left({U}^{(i+1)},\bm{U}^{(i+1)}\right) 
\mbox{  gives finite series}, &\mbox{then } \left({U}^{(i)},\bm{U}^{(i)}\right) \mbox{ gives infinite series,}
\end{array}
 \end{eqnarray}
 where $i=1,2,3,4$. This is true also for the hyperbolic potentials.
 
 To search finite- and infinite-series solutions we have to examine the coefficients $\gamma_n^{(i)}=\bm\gamma_n^{(i)}$ and use (\ref{truncation}).
 In general we need to assign values to $\ell$, $\gamma$ and $\delta$ as we see from 
 \begin{eqnarray}\label{gamas}
 \begin{array}{ll}
 \gamma_n^{(1)}=
2\beta\left(n-\ell-1\right),&\quad
\gamma_n^{(2)}=
2\beta\left(n-\ell-2\delta\right),\vspace{2mm}\\
\gamma_n^{(3)}=
2\beta\left(n+1-\ell-2\delta-2\gamma\right),&\quad
\gamma_n^{(4)}=
2\beta\left(n-\ell-2\gamma\right),\vspace{2mm}\\
 \gamma_n^{(5)}=
-2\beta\left(n-1+\ell+2\gamma+2\delta\right),& \quad
\gamma_n^{(6)}=
-2\beta\left(n+\ell+2\gamma\right),\vspace{2mm}\\
\gamma_n^{(7)}=
-2\beta\left(n+\ell+1\right),&\quad
\gamma_n^{(8)}=-2\beta\left(n+\ell+2\delta\right).
\end{array}
 \end{eqnarray}
Ushveridze claimed that there are finite-series solutions  when $\ell$ is a non-negative integer but  $\gamma_n^{(1)}$ and $\gamma_n^{(7)}$ imply finite series when $\ell$ is any  integer,  excepting $\ell=-1$. 
 
 On the other side,
 %
 %
 %
 if  $(\gamma,\delta)$ are given by
\begin{eqnarray}\label{4casos}
 \left(\gamma,\delta\right)=\left(\frac{1}{4},\frac{1}{4}\right), \ 
\left(\frac{1}{4},\frac{3}{4}\right), \ 
\left(\frac{3}{4},\frac{1}{4}\right) \mbox{ and }
 \left(\frac{3}{4},\frac{3}{4}\right),
\end{eqnarray}
 Eq. (\ref{schr-2a}) reduces to the Whittaker-Hill equation (WHE) -- 
 also called Hill's equation with three terms \cite{arscott,arscott1,arscott2} --
  \begin{eqnarray}
  \label{whittaher-hill}
\frac{d^2\psi}{du^2}+
\left\{ {\cal E}+
  4\beta^{2}{\sin^{4}u}- 4\beta\big[\beta+
 2(\gamma+\delta+\ell) \big]{\sin^{2}u}
  \right\}
  \psi=0.
\end{eqnarray}
%
 %
%
%
For this equation, there are finite series when $\ell$ is integer or half an odd integer, excepting for $i\eta=- \ell-\gamma-\delta=0$ which represents cases ruled by a Mathieu equation  that does not admit finite-series solutions. A similar WHE occurs also for the trigonometric Razavy potential \cite{razavy-trig}, 
%
%
\begin{eqnarray}
\label{razavy-trigonometrico}
 \mathcal{V}_r(u)&=&\frac{m^2}{8}\xi^2\left[1- \cos(2mu)\right]-m^2\xi(\bm{n}+1)\cos(mu)\nonumber\\
 &=&-m^2\left[(\bm{n}+1)\xi
+\xi^{2}\sin^{4}\frac{mu}{2}-\xi\big(\xi+
2+2\bm{n}\big)\sin^{2}\frac{mu}{2}\right], \quad [\text{Razavy potential}]
\end{eqnarray}
where $m\neq0$, $\bm{n}$  and $\xi$ are real parameters. For $\bm{n}=-1$ the Schr\"{o}dinger 
equation is solvable in terms of solutions of  Mathieu equation  \cite{razavy-trig,1932}.  Both, the
WHE  and the Mathieu equation, can as well be treated as a double confluent Heun equation
\cite{lea}.






\subsection{A  Whittaker-Hill equation   with $\delta=\gamma=1/4$ and Razavy potential}
%
 %
%
%
%
%
%
%
 %
 

 %
%
%
%
%
%
%
 %

As an example, we consider the case $\delta=\gamma=1/4$; after this
we comment on the solutions for the Razavy potential (\ref{razavy-trigonometrico}). 
 For this case
 \begin{eqnarray}
 \begin{array}{l}
  \psi(u)=U( z=\cos^2u),\quad 
z_0=1,\quad B_{1}=-\frac{1}{2}, \quad  B_{2}=1, \quad 
 B_{3}=-\frac{{\cal E}}{4}, \quad
i\omega=\beta,\quad i\eta= -\ell-\frac{1}{2}.
 \end{array}
 \end{eqnarray}
 Relations (\ref{gamas}) imply finite series if $\ell$  is an integer or half an odd integer ($\ell\neq -1/2$), 
but even for these $\ell$ there are bounded infinite-series expansions.

All the expansions  
$\left({U}^{(i)}, \bm{U}^{(i)}\right) $ give finite and infinite series solutions
$\left(\psi^{(i)}_{\;\ell}, \bm\psi^{(i)}_{\;\ell}\right) $. 
For infinite series, 
${\psi}_{\;\ell}^{(i)}$=  $\bm{\psi}_{\;\ell}^{(i)}$;
for finite series, ${\psi}_{\;\ell}^{(i)}$ and  
$\bm{\psi}_{\;\ell}^{(i)}$ are valid for opposite signs of $\beta$ due to the Arscott 
conditions (\ref{autovalores}). 
 With this proviso,
we can write only ${\psi}_{\;\ell}^{(i)}$.

From ${U}^{(1)}$, we find 
%
%
\begin{eqnarray}\label{b-1}
&&\begin{cases}
\psi_{\;\ell}^{(1)}(u)=e^{\beta z}\displaystyle \sum_{n}b_{n}^{(1)}
z^{n},\qquad\begin{cases}\mbox{if  } \ell=0,1,2,\cdots, \quad 0\leq n \leq \ell, \vspace{2mm}\\
\mbox{otherwise,}\quad 0\leq n< \infty,\quad\mbox{otherwise,},
\end{cases} \vspace{3mm}\\
\alpha_n^{(1)}=- \left(n+\frac{1}{2}\right)
\left(n+1\right), \quad \beta_n^{(1)}=
n\left(n-2\beta\right)-\frac{1}{4}{\cal E}+2\ell \beta+\frac{1}{2}\beta.\, \quad 
%
\gamma_n^{(1)}=
2\beta\left(n-\ell-1\right),
\end{cases}\vspace{3mm}\\
&&\hspace{1cm}
[
 \beta>0 \quad \mbox{for finite series},
\quad \beta\neq 0 \quad \mbox{for infinite series}
].\nonumber
\end{eqnarray}
In fact, $\alpha_{n-1}^{(1)}\gamma_n^{(1)}=- 2\beta\left(n-\frac{1}{2}\right)
n\left(n-\ell-1\right)>0$ assures the  Arscott conditions (\ref{autovalores})  
if $\beta>0$ ($1\leq n\leq\ell$). For $\beta<0$,  finite series satisfying  
conditions (\ref{autovalores}) are given by $\bm{\psi}_{\;\ell}^{(1)}$ as we
see from
%
\begin{eqnarray}\label{bb-1}
&&\begin{cases}
\bm{\psi}_{\;\ell}^{(1)}(u)=e^{\beta z}\displaystyle \sum_{n}\bm{b}_{n}^{(1)}
(z-1)^{n},\qquad\begin{cases}\mbox{if  } \ell=0,1,2,\cdots, \quad 0\leq n \leq \ell,\vspace{2mm}\\
\mbox{otherwise,}\quad 0\leq n< \infty,
\end{cases} \vspace{3mm}\\
\bm\alpha_n^{(1)}= \left(n+\frac{1}{2}\right)
\left(n+1\right), \quad \bm\beta_n^{(1)}=
n\left(n+2\beta\right)-\frac{1}{4}{\cal E}+
\frac{1}{2}\beta, \quad 
\gamma_n^{(1)}=
2\beta\left(n-\ell-1\right),
\end{cases}\vspace{3mm}\\
&&\hspace{1cm}
[
 \beta<0 \quad \mbox{for finite series},
\quad \beta\neq 0 \quad \mbox{for infinite series}].\nonumber
\end{eqnarray}
For one-term series ($\ell=0$), $\psi_{\ell=0}^{(1)}=\bm\psi_{\ell=0}^{(1)} $
with ${\cal E}=2\beta $. The two-term finite series ($\ell=1$)  satisfy the 
characteristic equation $\beta_{0}\beta_{1}- \alpha_{0}\gamma_{1} = 0 $
%
%
%
which gives
\begin{eqnarray*}
\begin{array}{l}
{\cal E}=2(3\beta+1)\pm 2\sqrt{4\beta^2+12\beta+1}\quad \mbox{for }\psi^{(1)};
\quad
{\cal E}=2(3\beta+1)\pm 2\sqrt{4\beta^2+1}\quad \mbox{for }\bm\psi^{(1)}.
\end{array}
\end{eqnarray*}
Thus, for $\psi^{(1)}$ the values of ${\cal E}$ are real if $\beta$ is positive 
as required by the Arscott theorem;  however, for $\bm\psi^{(1)}$,   ${\cal E}$ is real 
for positive and negative $\beta$,  despite the theorem.
%

%
%
From the other expansions $(U^{(i)},\bm{ U}^{(i)})$ we get additional 
eigenfunctions, $(\psi_{\;\ell}^{(i)},\bm{ \psi}_{\;\ell}^{(i)})$. Thus,
%
\begin{eqnarray}
&&\begin{cases}
\psi_{\;\ell}^{(2)}(u)=e^{\beta z}\;z^{\frac{1}{2}}\displaystyle \sum_{n}b_{n}^{(2)}
z^ n,\qquad\begin{cases}\mbox{if  } \ell=\frac{1}{2},\frac{3}{2},\frac{5}{2},\cdots, \quad 0\leq n \leq \ell-\frac{1}{2},
\vspace{2mm}\\
\mbox{otherwise,}\quad 0\leq n< \infty,
\end{cases}\vspace{2mm}\\
\alpha_n^{(2)}=- \left(n+\frac{3}{2}\right)
\left(n+1\right), \quad 
\beta_n^{(2)}=
n\left(n+1-2\beta\right)-\frac{1}{4}{\cal E}+2\beta\ell,\quad
\gamma_n^{(2)}=
2\beta\left(n-\ell-\frac{1}{2}\right),
\end{cases}\vspace{3mm}\\
&&\hspace{1cm}
[
 \beta>0 \quad \mbox{for finite series},
\quad \beta\neq 0 \quad \mbox{for infinite series}].\nonumber
\end{eqnarray}
Finite series for $\beta<0$ are given by $\bm{\psi}^{(2)}$. 
%
%
%
%
%
%
%
%
%
%
%
%
\begin{eqnarray}
&&\begin{cases}
\psi_{\;\ell}^{(3)}(u)=e^{\beta z}z^{\frac{1}{2}}(z-1)^{\frac{1}{2}}
\displaystyle \sum_{n}b_{n}^{(3)}
z^{n},\qquad\begin{cases}\mbox{if  } \ell=1,2,3,\cdots, \quad 0\leq n \leq \ell-1,\vspace{2mm}\\
\mbox{otherwise,}\quad 0\leq n< \infty,
\end{cases} \vspace{3mm}
\\
\alpha_n^{(3)}=-\left(n+\frac{3}{2}\right)
\left(n+1\right), \quad \beta_n^{(3)}=
n\left(n+2-2\beta\right)-\frac{1}{4}{\cal E}+1+2\beta\ell,\quad
 \gamma_n^{(3)}=
2\beta\left(n-\ell\right),
\end{cases}\vspace{3mm}\\
&&\hspace{1cm}
[
 \beta>0 \quad \mbox{for finite series},
\quad \beta\neq 0 \quad \mbox{for infinite series}].\nonumber
\end{eqnarray}
For $\beta<0$ finite series  are given by $\bm{\psi}^{(3)}$.
%
%
%
%
%
%
%
%
%
%
%
%
%
%
%
\begin{eqnarray}
&&\begin{cases}
\psi_{\;\ell}^{(4)}(u)=e^{\beta z}(z-1)^{\frac{1}{2}}
\displaystyle \sum_{n}b_{n}^{(4)}
z^{n}, \qquad\begin{cases}\mbox{if  } \ell=\frac{1}{2},\frac{3}{2},\frac{5}{2},\cdots, \quad 0\leq n \leq \ell-\frac{1}{2},
\vspace{2mm}\\
\mbox{otherwise,}\quad 0\leq n< \infty,
\end{cases}\\
\alpha_n^{(4)}=- \left(n+\frac{1}{2}\right)
\left(n+1\right), \quad \beta_n^{(4)}=
n\left(n+1-2\beta\right)-\frac{1}{4}{\cal E}+\frac{1}{4}+2\beta\ell+\frac{1}{2}\beta,\vspace{2mm}\\
   \gamma_n^{(4)}=
2\beta\left(n-\ell-\frac{1}{2}\right),
\end{cases}\vspace{3mm}  \\
&&\hspace{1cm}
[
 \beta>0 \quad \mbox{for finite series},
\quad \beta\neq 0 \quad \mbox{for infinite series}].\nonumber
\end{eqnarray}
$\bm{\psi}^{(4)}$ 
gives finite series for $\beta<0$.
%
%
%
%
%
%
%
%
%
%
%
%
%
%
%
%
%
%
\begin{eqnarray}
&&\begin{cases}
\psi_{\;\ell}^{(5)}(u)=e^{-\beta z}\displaystyle \sum_{n}b_{n}^{(5)}
z^{n},\qquad\begin{cases}\mbox{if  } \ell=-1,-2,-3,\cdots, \quad 0\leq n \leq -\ell-1,\vspace{2mm}\\
\mbox{otherwise,}\quad 0\leq n< \infty,
\end{cases} \vspace{2mm}\\
\alpha_n^{(5)}=- \left(n+\frac{1}{2}\right)
\left(n+1\right), \quad \beta_n^{(5)}=
n\left(n+2\beta\right)-\frac{1}{4}{\cal E}+2\beta\ell+\frac{3}{2}\beta,\quad
    \gamma_n^{(5)}=-
2\beta\left(n+\ell\right),
\end{cases}\vspace{3mm}\\
&&\hspace{1cm}
[
 \beta<0 \quad \mbox{for finite series},
\quad \beta\neq 0 \quad \mbox{for infinite series}
].\nonumber
\end{eqnarray}
Finite series for $\beta>0$ result from the expansions $\bm{\psi}^{(5)}$.
%
%
%
%
%
%
%
%
%
%
%
%
%
%
\begin{eqnarray}
&&\begin{cases}
\psi_{\;\ell}^{(6)}(u)=e^{-\beta z}\;z^{\frac{1}{2}}\displaystyle \sum_{n}b_{n}^{(6)}
z^{n},
\qquad
\begin{cases}\mbox{if  } \ell=-\frac{3}{2},-\frac{5}{2},-\frac{7}{2},\cdots, \quad 0\leq n \leq -\ell-\frac{3}{2},\vspace{2mm}\\
\mbox{otherwise,}\quad 0\leq n< \infty,
\end{cases}
\vspace{2mm}\\
\alpha_n^{(6)}=- \left(n+\frac{3}{2}\right)
\left(n+1\right), \quad 
\beta_n^{(6)}=
n\left(n+1+2\beta\right)-\frac{1}{4}{\cal E}+\frac{1}{4}+2\beta\ell+\frac{5}{2}\beta, \vspace{2mm}\\
\gamma_n^{(6)}=-
2\beta\left(n+\ell+\frac{1}{2}\right),
\end{cases}\vspace{3mm}\\
&&\hspace{1cm}
[
 \beta<0 \quad \mbox{for finite series},
\quad \beta\neq 0 \quad \mbox{for infinite series}].\nonumber
\end{eqnarray}
Finite-series solutions for $\beta>0$ come from $\bm{\psi}^{(6)}$.
%
%
%
%
%
%
%
%
%
%
%
%
%
%
%
\begin{eqnarray}
&&\begin{cases}
\psi_{\;\ell}^{(7)}(u)=e^{-\beta z}z^{\frac{1}{2}}
(z-1)^{\frac{1}{2}}\displaystyle \sum_{n}b_{n}^{(7)}
z^{n},\quad\begin{cases}\mbox{if  } \ell=-2,-3,-4,\cdots, \quad
%
0\leq n \leq -\ell-2,\vspace{2mm}\\
\mbox{otherwise,}\quad 0\leq n< \infty,
\end{cases}\vspace{2mm}\\
\alpha_n^{(7)}=- \left(n+\frac{3}{2}\right)
\left(n+1\right), \quad \beta_n^{(7)}=
n\left(n+2 +2\beta\right)=\frac{1}{4}{\cal E}+1+2\beta\ell+\frac{5}{2}\beta,\vspace{2mm}\\
\gamma_n^{(7)}=
-2\beta\left(n+\ell+1\right),
\end{cases}\vspace{3mm}\\
&&\hspace{1cm}
[
 \beta<0 \quad \mbox{for finite series},
\quad \beta\neq 0 \quad \mbox{for infinite series}
].\nonumber
\end{eqnarray}
Finite series for $\beta>0$ result from the expansions $\bm{\psi}^{(7)}$.  
%
%
%
%
%
%
%
%
%
%
%
%
%
%
%
%
\begin{eqnarray}\label{b-8}
&&\begin{cases}
\psi_{\;\ell}^{(8)}(u)=e^{-\beta z}(z-1)^{\frac{1}{2}}
\displaystyle \sum_{n}b_{n}^{(8)}
z^{n},\quad\begin{cases}\mbox{if  } \ell=-\frac{3}{2},-\frac{5}{2},-\frac{7}{2},\cdots, \quad
%
 0\leq n \leq -\ell-\frac{3}{2},\vspace{2mm}\\
\mbox{otherwise,}\quad 0\leq n< \infty,
\end{cases}
\vspace{3mm}\\
\alpha_n^{(8)}=- \left(n+\frac{1}{2}\right)
\left(n+1\right), \qquad \beta_n^{(8)}=
n\left(n+1+2\beta\right)-\frac{1}{4}{\cal E}+\frac{1}{4}+2\beta\ell+\frac{3}{2}\beta,\vspace{2mm}\\
 \gamma_n^{(8)}=
-2\beta\left(n+\ell+\frac{1}{2}\right),
\end{cases}\vspace{3mm}\\
&&\hspace{1cm}
[
 \beta<0 \quad \mbox{for finite series},
\quad \beta\neq 0 \quad \mbox{for infinite series}
].\nonumber
\end{eqnarray}
For $\beta>0$ finite series result from the expansions $\bm{\psi}^{(8)}$.
%
%
%
%
%
%
%
%
%
%
%
%
%
%



%

Then, if $\ell$  is an integer or half an odd integer ($\ell\neq -1/2$), the 16 expansions lead to bounded finite-series expansions,  valid for different ranges of the
parameters $\beta$ or $\ell$; however,  there are also infinite series solutions
which include the ones specified in (\ref{prescription}).
%
%
From the solutions (\ref{b-1}-\ref{b-8}) combined with relations (\ref{a1}-\ref{a3})
for infinite series, we find the following solutions.
\begin{align}
&\mbox{If }\ell=0: \begin{cases}
\mbox{finite series: } \psi_{\;\ell}^{(1)}=\bm\psi_{\;\ell}^{(1)}
,\vspace{2mm}\\
\mbox{infinite series: }   \psi_{\;\ell}^{(2)}=\psi_{\;\ell}^{(3)}=\psi_{\;\ell}^{(6)}=\psi_{\;\ell}^{(7)}, \quad \psi_{\;\ell}^{(4)}=\psi_{\;\ell}^{(5)}=\psi_{\;\ell}^{(8)}  ,
\end{cases}\vspace{3mm}\\
&\mbox{If }\ell=1,2,3,\cdots: \begin{cases}
\mbox{finite series: } \psi_{\;\ell}^{(1)} \mbox{ and }\psi_{\;\ell}^{(3)} \mbox{ if } \beta>0,
\quad  \bm\psi_{\;\ell}^{(1)} \mbox{ and }\bm\psi_{\;\ell}^{(3)} \mbox{ if } \beta<0,
\vspace{2mm}\\
\mbox{infinite series: }   \psi_{\;\ell}^{(2)}=\psi_{\;\ell}^{(6)}=\psi_{\;\ell}^{(7)}, \quad \psi_{\;\ell}^{(4)}=\psi_{\;\ell}^{(5)}=\psi_{\;\ell}^{(8)} .
\end{cases}\vspace{4mm}\\
&\mbox{If }\ell=\frac{1}{2},\frac{3}{2},\frac{5}{2},\cdots: \begin{cases}
\mbox{finite series: } \psi_{\;\ell}^{(2)} \mbox{ and }\psi_{\;\ell}^{(4)} \mbox{ if } \beta>0,
\quad  \bm\psi_{\ell}^{(2)} \mbox{ and }\bm\psi_{\ell}^{(4)} \mbox{ if } \beta<0,
\vspace{2mm}\\
\mbox{infinite series: }   \psi_{\;\ell}^{(1)}=\psi_{\;\ell}^{(5)}=\psi_{\;\ell}^{(8)}, \quad \psi_{\;\ell}^{(3)}=\psi_{\;\ell}^{(6)}=\psi_{\;\ell}^{(8)} .
\end{cases}\end{align}
\begin{align}
&\mbox{If }\ell=-1 :\begin{cases}
\mbox{finite series: } \psi_{\;\ell}^{(5)} = \bm\psi_{\ell}^{(5)} ,
\vspace{2mm}\\
\mbox{infinite series: }   \psi_{\;\ell}^{(1)}=\psi_{\;\ell}^{(4)}=\psi_{\;\ell}^{(8)}, \quad \psi_{\;\ell}^{(2)}=\psi_{\;\ell}^{(3)}=\psi_{\;\ell}^{(6)}= \psi_{\;\ell}^{(7)}.
\end{cases}\vspace{4mm}\\
&\mbox{If }\ell=-2,-3,-4,\cdots: \begin{cases}
\mbox{finite series: } \psi_{\;\ell}^{(5)} \mbox{ and }\psi_{\;\ell}^{(7)} \mbox{ if } \beta<0,
\quad  \bm\psi_{\ell}^{(5)} \mbox{ and }\bm\psi_{\ell}^{(7)} \mbox{ if } \beta>0,
\vspace{2mm}\\
\mbox{infinite series: }   \psi_{\;\ell}^{(1)}=\psi_{\;\ell}^{(4)}=\psi_{\;\ell}^{(8)}, \quad \psi_{\;\ell}^{(2)}=\psi_{\;\ell}^{(3)}=\psi_{\;\ell}^{(6)} .
\end{cases}\vspace{4mm}\\
%
&\mbox{If }\ell=-\frac{3}{2},-\frac{5}{2},-\frac{7}{2},\cdots: \begin{cases}
\mbox{finite series: } \psi_{\;\ell}^{(6)} \mbox{ and }\psi_{\;\ell}^{(8)} \mbox{ if } \beta<0,
\quad  \bm\psi_{\ell}^{(6)} \mbox{ and }\bm\psi_{\ell}^{(8)} \mbox{ if } \beta>0,
\vspace{2mm}\\
\mbox{infinite series: }   \psi_{\;\ell}^{(1)}=\psi_{\;\ell}^{(4)}=\psi_{\;\ell}^{(5)}, \quad \psi_{\;\ell}^{(2)}=\psi_{\;\ell}^{(3)}=\psi_{\;\ell}^{(7)} .
\end{cases}
\end{align}


So, 
the previous finite-series solutions  correspond to  two  infinite-series solutions for the Whittaker-Hill equation. In fact, for such equation,   there are two bounded infinite-series expansions for any real value of $\ell$ 
since by (\ref{a4})
\begin{eqnarray}
&&	\mbox{if } \ell\mbox{  is neither  an integer nor half an odd integer}, 
	\mbox{ then}\nonumber\\
&&	{\psi}_{\;\ell}^{(1)}= {\psi}_{\;\ell}^{(4)}=
	{\psi}_{\;\ell}^{(5)}= {\psi}_{\;\ell}^{(8)}
	\mbox{ and }{\psi}_{\;\ell}^{(2)}= {\psi}_{\;\ell}^{(3)}=
	{\psi}_{\;\ell}^{(6)}= {\psi}_{\;\ell}^{(7)} \mbox{ are infinite series solutions.} 
\end{eqnarray}

Therefore, for any real value of $\ell$ there are two infinite-series expansions, regardless of finite-series solutions. The same holds for the Razavy potential (\ref{razavy-trigonometrico}) with  $\bm{n}$ taking the role o $\ell$. 
In effect, the substitutions
%
%
\begin{eqnarray}\begin{array}{l}
z= \cos^{2}\left(\frac{mu}{2}\right),\qquad \psi(u)=U(z),\qquad [\text{ for Razavy potential}]
\end{array}
\end{eqnarray}
transform the  Schr\"{o}dinger equation in
%
%
%
%
%
 \begin{eqnarray}
&&\begin{array}{l}
z(z-1)\frac{d^{2}U}{dz^{2}}+\left(z-\frac{1}{2}\right)
\frac{dU}{dz}-\Big[(\bm{n}+1)\xi+\frac{{\cal E}}{m^2}
+2(\bm{n}+1)\xi(z-1)
+\xi^2z(z-1)\Big]U=0,\quad
\end{array} 
\end{eqnarray}
which is a CHE with
 \begin{eqnarray}\label{trig-r}
\begin{array}{l}
z_0=1,\quad B_{1}=-\frac{1}{2}, \quad  B_{2}=1, \quad 
 B_{3}=-(\bm{n}+1)\xi-\frac{{\cal E}}{m^2},\quad 
%
i\omega=\xi,\quad i\eta= -\bm{n}-1.
 \end{array}
 \end{eqnarray}
 %
%
	Hence
	\begin{eqnarray*}
 \begin{array}{llll}					
 \gamma_n^{(1)}=\xi\left(n- \bm{n}-\frac{3}{2}\right),& \gamma_n^{(2)}= \xi\left(n- \bm{n}-1\right),&
  \gamma_n^{(3)}=\xi\left(n- \bm{n}-\frac{1}{2}\right),&\gamma_n^{(4)}=\xi\left(n- \bm{n}-1\right), 
  \vspace{2mm}\\
 \gamma_n^{(5)}=-\xi\left(n+ \bm{n}+\frac{1}{2}\right),& \gamma_n^{(6)}= -\xi\left(n+ \bm{n}+1\right),& \gamma_n^{(7)}=-\xi\left(n+ \bm{n}+\frac{3}{2}\right),& \gamma_n^{(8)}=-\xi\left(n+ \bm{n}+1\right).
\end{array}
\end{eqnarray*}
%
	 %
%
%
%
%
%
%
 %
 A particular consequence of this is the occurrence of finite series if $\bm{n}\neq-1$  is an integer or half an odd integer. 
 
Finite-series solutions for a modified version of the Razavy potential have been considered in
Ref. \cite{dong-2018}  when  $\bm{n}$ is a non-negative integer.

 \subsection{Other cases}

Although we need the values of the parameters 
to find finite- and infinite-series solutions, 
 for two cases it is sufficient 
 to specify lower or upper values for $\gamma$ and $\delta$. In effect, by using  (\ref{trig-a}), we find that 
  $\gamma_{n}^{(1)}=\bm\gamma_{n}^{(1)}= 2\beta(n-\ell-1)$ and 
  $\gamma_{n}^{(7)}=\bm\gamma_{n}^{(7)}= -2\beta(n+\ell+1)$ do not depend on $\gamma$ and $\delta$; then, 
  from  (\ref{trig-1})  
  we get solutions $(\psi_{\;\ell}^{(1)}, 
  \bm\psi_{\;\ell}^{(1)})$ and $(\psi_{\;\ell}^{(7)}, 
  \bm\psi_{\;\ell}^{(7)})$ admitting finite series when $\ell\neq -1$ is integer.
 %
 %
%
 %
 %
%
\begin{eqnarray}\label{p-1}
&&\begin{cases}
\psi_{\;\ell}^{(1)}(z)=e^{\beta z}  z^{\delta-\frac{1}{4}}(z-1)^{\gamma-\frac{1}{4}}\displaystyle \sum_{n}b_{n}^{(1)}
z^{n},\qquad  \begin{cases}\mbox{if  } \ell=0,1,2,\cdots, \quad 0\leq n \leq \ell,\vspace{2mm}  \\
\mbox{otherwise,}\quad 0\leq n< \infty,
\end{cases} \vspace{3mm}\\
\alpha_n^{(1)}=- \left(n+2\delta\right)
\left(n+1\right), \qquad \beta_n^{(1)}=
n\left(n+2\gamma+2\delta-1-2\beta\right)-\frac{1}{4}{\cal E}+\left(\gamma+\delta-\frac{1}{2}\right)^2\\
%
+2\beta(\ell+\gamma),\qquad \gamma_n^{(1)}=
2\beta\left(n-\ell-1\right),
\end{cases}\vspace{2mm}\\
&&\hspace{.5cm}\begin{array}{l}\left[
\gamma\geq\frac{1}{4},\quad\delta\geq\frac{1}{4}:\quad \beta>0\quad \mbox{for finite series},\quad
\beta\neq 0\quad \mbox{for infinite series}\right];\end{array}\nonumber
\end{eqnarray}
$\bm{\psi}_{\;\ell}^{(1)}$ gives finite series for $\beta<0$.
%
Infinite series valid  for  $\ell=0,1,2.\cdots$ come from 
   $U^{(5)}$ or  $\bm{U}^{(5)}$, 
  \begin{eqnarray}\label{psi-5}
{\psi}_{\;\ell}^{(5)}(z)=e^{-\beta z}  z^{\delta-\frac{1}{4}}(z-1)^{\gamma-\frac{1}{4}}
\displaystyle \sum_{n=0}^{\infty}b_{n}^{(5)}z^{n}=\bm{\psi}_{\;\ell}^{(5)}(z),\quad \gamma\geq\frac{1}{4},
\quad \delta\geq\frac{1}{4},\quad \ell=0,1,\cdots,
%
%
%
\end{eqnarray}
where $\gamma_n^{(5)}=
-2\beta\left(n+\ell+2\delta+2\gamma-1\right)\neq 0$.
Therefore, if $\delta\geq\frac{1}{4}$ and  $\gamma\geq\frac{1}{4}$, there are infinite series  for any real 
value of $\ell$. On the other side,
  \begin{eqnarray}
  &&\begin{cases}
  \psi_{\;\ell}^{(7)}(z)=e^{-\beta z} z^{\frac{3}{4}-\delta}(z-1)^{\frac{3}{4}-\gamma} \displaystyle \sum_{n}b_{n}^{(7)}
  z^{n},\;\begin{cases}\mbox{if  } \ell=-2,-3,-4,\cdots,\;
%
0\leq n \leq -\ell-2,\vspace{2mm}\\
\mbox{otherwise,}\quad 0\leq n< \infty,
\end{cases}
\vspace{2mm}\\
  \alpha_n^{(7)}=- \left(n+2-2\delta\right)
  \left(n+1\right), \quad \beta_n^{(7)}=
  n\left(n+3-2\gamma-2\delta+2\beta\right)-\frac{1}{4}{\cal E}+\left(\gamma+\delta-\frac{1}{2}\right)^2 \\
 + 2-2\gamma-2\delta+2\beta(\ell+\gamma+1),\qquad \gamma_n^{(7)}=
  -2\beta\left(n+\ell+1\right),
  \end{cases}\vspace{3mm}\\
&&\hspace{.5cm}\begin{array}{l}\left[
\gamma\leq\frac{3}{4},\quad\delta\leq\frac{3}{4}:\quad \beta<0\quad \mbox{for finite series},\quad
\beta\neq 0\quad \mbox{for infinite series}\right],\end{array}\nonumber
  \end{eqnarray}
  whereas $\bm{\psi}_{\;\ell}^{(7)}$ gives finite series for $\beta>0$. 
Infinite series valid also for $\ell=-2,-3,-4,\cdots$  come from 
$U^{(3)} $ or $\bm{U}^{(3)}$. In fact,
\begin{eqnarray}\label{psi-3}
\psi_{\;\ell}^{(3)}(z)=e^{\beta z} z^{\frac{3}{4}-\delta}(z-1)^{\frac{3}{4}-\gamma} \displaystyle \sum_{n=0}^{\infty}b_{n}^{(3)}
z^{n}=\bm\psi_{\;\ell}^{(3)}(z),\begin{array}{l} \quad  \gamma\leq\frac{3}{4},\quad\delta\leqslant\frac{3}{4},\quad \ell= -2,-3,\cdots,\end{array} 
%
%
%
\end{eqnarray}
where $\gamma_n^{(3)}=
2\beta\left(n+1-\ell -2\delta-2\gamma\right)\neq 0$.
Therefore,   if $\delta\leq\frac{3}{4}$ and  $\gamma\leq\frac{3}{4}$, there are infinite series  for any real 
value of $\ell$.


 %

The previous expansions include three different cases: { (1)  $\delta<\frac{1}{4}$ and  $\gamma<\frac{1}{4}$; (2)  $\delta>\frac{3}{4}$ and  $\gamma>\frac{3}{4}$}; (3)  
$\frac{1}{4}\leq\delta\leq\frac{3}{4}$ and  $\frac{1}{4}\leq\gamma\leq\frac{3}{4}$.

							
{ (1) For $\delta<\frac{1}{4}$ and  $\gamma<\frac{1}{4}$,  
$\left(  \psi_{\;\ell}^{(7)}, \bm\psi_{\;\ell}^{(7)}\right)$ 
represent  finite-series solutions  for $\ell=-2,-3,-4,\cdots$; there is one infinite-series expansion for each $\ell$: 
 $  \psi_{\;\ell}^{(7)}=\bm\psi_{\;\ell}^{(7)}$ if  $\ell\neq -2,-3,-4,\cdots$ 
 and 
 $  \psi_{\;\ell}^{(3)}$ for $\ell\leq1$  [see remarks following relations (\ref{psi-3})]}. 


 %
 %
%
%

 								
 %
{(2) For $\delta>\frac{3}{4}$ and  $\gamma>\frac{3}{4}$, $\left(  \psi_{\;\ell}^{(1)}, \bm\psi_{\;\ell}^{(1)}\right)$ represent finite-series solutions  only for non-negative  integer values of $\ell$, in opposition to the first case;  
 there is one infinite-series expansion for each $\ell$: 
 $  \psi_{\;\ell}^{(1)}=\bm\psi_{\;\ell}^{(1)}$ if  $\ell\neq 0,1,2,\cdots$ 
 and 
 $  \psi_{\;\ell}^{(5)}$ for $\ell\geq-2$  [see remarks following relations (\ref{psi-5})]}. 
%
%
%
 %
 %
 %
%

 (3)  For $\frac{1}{4}\leq\delta\leq\frac{3}{4}$ and  $\frac{1}{4}\leq\gamma\leq\frac{3}{4}$, as in the cases of the Whittaker-Hill equation,  there are finite series for positive e negative
 integer values of $\ell$:  $(\psi_{\;\ell}^{(1)}, 
  \bm\psi_{\;\ell}^{(1)})$ and $(\psi_{\;\ell}^{(7)}, 
  \bm\psi_{\;\ell}^{(7)})$ for non-negative and negative integer values of
$\ell$, respectively, excepting $\ell=-1$. There are also  infinite-series expansions  for  any real value of $\ell$.
%
 %
%
%
%
%
%
%
%

 %
 %

%
%

Note that, according to (\ref{gamma-delta}),  there are bounded expansions for other ranges of
$\gamma$ and $\delta$, but the nature of the expansions depend on the  values of the parameters.

 %
%
%
%
%
%
%


\section{The trigonometric potential given in Eq.  (\ref{ush-2-trigonometrico})}

For the trigonometric potential (\ref{ush-2-trigonometrico}),
the Schr\"{o}dinger equation (\ref{schr}) becomes
  \begin{eqnarray}
&&
\frac{d^2\psi}{du^2}+
\big[{\cal E}+\begin{array}{l}4\gamma^{2}\cos^{4}u-4\gamma\left(\gamma+2-2\beta \right)\cos^{2}u-4\left( \delta-\frac{1}{4}\right)
\left(\delta-\frac{3}{4}\right)\cot^2u\end{array}\nonumber\vspace{3mm}\\
&&
-\begin{array}{l}4\left( \beta+\delta+\ell-\frac{1}{4}\right)\left(\beta+\delta+\ell-\frac{3}{4}\right)\tan^2u\big]\psi=0\end{array}.
\end{eqnarray}
The substitutions
 \begin{eqnarray}\label{trig-2}
  z=\cos^2u,\qquad \psi(u)=
  z^{\beta+\delta+\ell-\frac{1}{4}}(z-1)^{\delta-\frac{1}{4}}U(z),
  \qquad\left[0\leq z\leq 1\right] 
   \end{eqnarray}
 transform the above equation into a confluent Heun equation,
 \begin{eqnarray*}
&&\begin{array}{l}
z(z-1)\frac{d^{2}U}{dz^{2}}+[-2(\beta+\delta+\ell)+2(\beta+2\delta+\ell)z]
\frac{dU}{dz}+\Big[-\frac{{\cal E}}{4}+2\delta(\beta+\delta+\ell)+2\gamma(1-\beta)-\frac{1}{8}
\end{array}\vspace{2mm}\\
%
&&+2\gamma(1-\beta)(z-1)-\gamma^{2}z(z-1)\Big]U=0.
\end{eqnarray*}
 Thence the parameters of the CHE (\ref{che}) are
 \begin{eqnarray}
&& \begin{array}{l}
z_0=1,\qquad B_{1}=-2(\beta+\delta+\ell), \qquad  B_{2}=2(\beta+2\delta+\ell), \end{array} \quad
 \vspace{2mm}\nonumber\\
&& 
 B_{3}=-\frac{{\cal E}}{4}+2\delta(\beta+\delta+\ell)+2\gamma(1-\beta)-\frac{1}{8},\qquad i\omega=-\gamma,\quad i\eta= \beta-1,
 \end{eqnarray}
 %

 The wavefunctions (\ref{trig-2})  corresponding to  $\left({U}_{\;\ell}^{(i)},\bm{U}_{\;\ell}^{(i)}\right)$ 
  are again denoted by $\left({\psi}_{\;\ell}^{(i)},\bm{\psi}_{\;\ell}^{(i)}\right)$ and become  
  bounded for any value of $z$  by imposing constraints on 
  $\delta$ and $\ell+\beta+\delta$, namely,
  %
  %
\begin{eqnarray} \label{gamma-delta-2}
\begin{array}{l}
\left({\psi}_{\;\ell}^{(1)},\bm{\psi}_{\;\ell}^{(1)}\right) ,\left({\psi}_{\;\ell}^{(5)},\bm{\psi}_{\;\ell}^{(5)}\right): \ell+\beta+\delta\geq\frac{1}{4},\quad\delta\geq\frac{1}{4};\vspace{3mm}\\
\left({\psi}_{\;\ell}^{(2)},\bm{\psi}_{\;\ell}^{(2)}\right) ,\left({\psi}_{\;\ell}^{(6)},\bm{\psi}_{\;\ell}^{(6)}\right): \ell+\beta+\delta\leq\frac{3}{4}, 
\quad\delta\geq\frac{1}{4};\vspace{3mm}\\
\left({\psi}_{\;\ell}^{(3)},\bm{\psi}_{\;\ell}^{(3)}\right),\left({\psi}_{\;\ell}^{(7)},\bm{\psi}_{\;\ell}^{(7)}\right):  \ell+\beta+\delta\leq\frac{3}{4}, \quad\delta\leq\frac{3}{4};\vspace{3mm}\\
\left({\psi}_{\;\ell}^{(4)},\bm{\psi}_{\;\ell}^{(4)}\right),\left({\psi}_{\;\ell}^{(8)},\bm{\psi}_{\;\ell}^{(8)}\right):  \ell+\beta+\delta\geq\frac{1}{4},  \quad\delta\leq\frac{3}{4}.
\end{array}
\end{eqnarray}
These restrictions  assure that the exponents of  $z$ and $z-1$, respectively, are not negative 
numbers; consequently the solutions are well behaved when $z=0$ and $z=1$.  Conditions 
(\ref{gamma-delta-2}) depend on $\ell$
in contrast  with conditions (\ref{gamma-delta})  for the previous potential.
 Moreover, the two pairs of solutions collected above make possible to find, at the same time, 
 solutions in finite and infinite series for the same range of  $\ell+\beta+\delta $ and $\delta$.
 This is consistent with (\ref{prescription}).

 As before, to get finite- and infinite-series solutions we inspect  $\gamma_n^{(i)}=\bm\gamma_n^{(i)}$:
 \begin{eqnarray}\begin{array}{ll}
 \gamma_n^{(1)}=-2\gamma(n+2\beta+2\delta+\ell-2),&\quad
 \gamma_n^{(2)}=-2\gamma(n-\ell-1),\vspace{2mm}\\
\gamma_n^{(3)}=-2\gamma(n-\ell-2\delta),&\quad
\gamma_n^{(4)}=-2\gamma(n+2\beta+\ell-1),\vspace{2mm}\\
%
 \gamma_n^{(5)}=2\gamma(n+ \ell+2\delta),&\quad
\gamma_n^{(6)}=
2\gamma(n-2\beta-\ell+1),\vspace{2mm}\\
\gamma_n^{(7)}=2\gamma(n-2\beta-
2\delta-\ell+2),&\quad
\gamma_n^{(8)}=2\gamma(n+\ell+1).
\end{array}
 \end{eqnarray}
Thus, $\gamma_{\;n}^{(2)}$ implies finite series if $\ell=0,1,2,\cdots$,  as stated by Ushveridze; 
however,
 $\gamma_{\;n}^{(8)}$ implies finite series also for $\ell=-2,-3,-4,\cdots$. Moreover, for $\delta=1/4$ 
 or $\delta=3/4$,
 $\gamma_{\;n}^{(3)}$ and $\gamma_{\;n}^{(5)}$ give finite series when $\ell$ 
 is half an odd integer, excepting $\ell=-1/2$ (if $\delta=1/4$) and  $\ell=-3/2$ (if $\delta=3/4$).



\subsection{Solutions for $\delta=1/4$ 
and possible extension for another potential}

%

As an illustration we discuss the case $\delta=1/4$, that is, 
 \begin{eqnarray}
&& \begin{array}{l}
\delta=\frac{1}{4},\quad z_0=1,\qquad B_{1}=-2(\beta+\ell)-\frac{1}{2}, \qquad  B_{2}=2(\beta+\ell)+1, \end{array} \quad
 \vspace{2mm}\nonumber\\
&& 
 B_{3}=-\frac{{\cal E}}{4}+2\delta(\beta+\delta+\ell)+2\gamma(1-\beta)-\frac{1}{8},\qquad i\omega=-\gamma,\quad i\eta= \beta-1.
 \end{eqnarray}
There are finite-series solutions when $\ell$ is integer and when $\ell$ is half an odd integer. The former are written in section 4.2.

The expansions $U^ {(3)}$ give
\begin{eqnarray}
\begin{cases}
\psi_{\;\ell}^{(3)}(z)=e^{-\gamma z}z^{-\beta-\ell+\frac{1}{2}}\;
(z-1)^{\frac{1}{2}}\displaystyle \sum_{n}b_{n}^{(3)}
z^{n},\qquad  \begin{cases}\mbox{if  } \ell=\frac{1}{2},\frac{3}{2},\frac{5}{2},\cdots, \quad 0\leq n \leq \ell-\frac{1}{2},\vspace{2mm}  \\
\mbox{otherwise,}\quad 0\leq n< \infty,
\end{cases}  \vspace{2mm}\\
\alpha_n^{(3)}=- \left[n-2(\beta+\ell)+\frac{3}{2}\right]
\left(n+1\right), \qquad \beta_n^{(3)}=
n\left(n+3-B_{2}-2i\omega z_{0}\right)+\\
2-B_2+B_{3}+2\eta\omega z_0- i\omega (B_{1}+2z_0),\qquad \gamma_n^{(3)}=
-2\gamma\left(n-\ell-\frac{1}{2}\right),\
\end{cases}\vspace{2mm}\\
\begin{array}{l}
\left[\delta=\frac{1}{4}, \quad \beta+\ell\leq\frac{1}{2}: \quad \gamma<0 \quad \mbox{for finite series},\quad
\gamma\neq 0 \quad \mbox{for infinite series} \right].\end{array}\nonumber
\end{eqnarray}
Finite series for $\gamma>0$ are obtained from $\bm{\psi}_{\;\ell}^{(3)}$
obtained from $\bm{U}^ {(3)}$.
Infinite series for $\ell=\frac{1}{2},\frac{3}{2},\cdots $ come
from $U^{(7)}$  and take the form
%
\begin{eqnarray}
\psi_{\;\ell}^{(7)}(z)=e^{\gamma z}z^{-\beta-\ell+\frac{1}{2}}\;
(z-1)^{\frac{1}{2}}\displaystyle \sum_{n=0}^ {\infty}b_{n}^{(7)}
z^{n},\qquad \begin{array}{l} \delta=\frac{1}{4}, \quad \beta+\ell\leq\frac{1}{2},
\quad \ell=\frac{1}{2},\frac{3}{2},\frac{5}{2},\cdots 
\end{array}
%
\end{eqnarray}
since $\gamma_n^{(7)}=2\gamma\left(n-2\beta-\ell+\frac{3}{2}\right)\neq 0$ because
$-2\beta\geq -1+2\ell$.
%


     The expansions $U^ {(5)}$ give finite series when $\ell=-\frac{3}{2},-\frac{5}{2},-\frac{7}{2},\cdots$,  namely,
\begin{eqnarray}
\begin{cases}
\psi_{\;\ell}^{(5)}(z)=e^{\gamma z}z^{\beta+\ell}
\displaystyle \sum_{n}b_{n}^{(5)}
z^{n},\qquad  \begin{cases}\mbox{if  } \ell=-\frac{3}{2}-\frac{5}{2},-\frac{7}{2},\cdots, 
\quad 0\leq n \leq -\ell-\frac{3}{2};\vspace{2mm}  \\
\mbox{otherwise,}\quad 0\leq n< \infty;
\end{cases}  \vspace{2mm}\\
\alpha_n^{(5)}=- \left[n+2(\beta+\ell)+\frac{3}{2}\right]
\left(n+1\right), \qquad \beta_n^{(5)}=
n\left(n+B_{2}-1+2i\omega z_{0}\right)+\\
%
B_{3}+2\eta\omega z_0- i\omega B_{1},\qquad 
\gamma_n^{(5)}=
2\gamma\left(n+\ell+\frac{1}{2}\right),
\end{cases}\vspace{2mm}\\
\begin{array}{l}
\left[\delta=\frac{1}{4}, \quad \beta+\ell\geq 0: \quad \gamma>0 \quad 
\mbox{for finite series},\quad
\gamma\neq 0 \quad \mbox{for infinite series} \right].\end{array}\nonumber
\end{eqnarray}
Finite series for $\gamma<0$ are obtained from $\bm{\psi}_{\;\ell}^{(5)}(z)$.
If $ \ell=-\frac{3}{2}-\frac{5}{2},\cdots$, infinite series come 
from 
$U^{(1)}$, 
\begin{eqnarray}
\psi_{\;\ell}^{(1)}(z)=e^{-\gamma z}z^{\beta+\ell}
\displaystyle \sum_{n=0}^{\infty}b_{n}^{(1)}
z^{n},\quad  \begin{array}{l}
\delta=\frac{1}{4}, \quad\beta+\ell\geq 0,\quad  \ell=-\frac{3}{2}-\frac{5}{2},-\frac{7}{2},\cdots,
\end{array} 
%
\end{eqnarray}
where $\gamma_n^{(1)}=
-2\gamma\left(n+2\beta+\ell-\frac{3}{2}\right)\neq 0 $ for the above parameters.

%
%

The previous discussion can be applied to a potential given in Ref. 
\cite{pozo}, namely,
\begin{eqnarray}\label{pozo}
\mathcal{V}(u) = -4\gamma^2 \cos^4u +4\gamma(2N + \gamma+\delta +1)\cos^2u +
\delta(\delta-1)\tan^2u,
\end{eqnarray}
where $\gamma$ and $\delta$ are real parameters. Also this admit finite-series solutions when $N$ is an integer
or half an odd integer (besides infinite-series solutions) as we see from the following expressions. The substitutions
 \begin{eqnarray}
  z=\cos^2u,\qquad \psi(u)=
  z^{\delta/2}U(z),
  \qquad\left[0\leq z\leq 1\right] 
   \end{eqnarray}
 transform the Schr\"odinger equation  into
 \begin{eqnarray*}
&&\begin{array}{l}
z(z-1)\frac{d^{2}U}{dz^{2}}+[-\delta-\frac{1}{2}+(1+\delta)z]
\frac{dU}{dz}+\Big[-\frac{{\cal E}-\delta}{4}+\gamma(1+\delta+2N)+\gamma(2N+\delta+1)(z-1)
\end{array}\vspace{2mm}\\
%
&&-\gamma^{2}z(z-1)\Big]U=0.
\end{eqnarray*}
which is the  CHE (\ref{che}) with $z_0=1$ and 
 \begin{eqnarray}
\begin{array}{l}
 B_{1}=-\delta-\frac{1}{2}, \quad  B_{2}=1+\delta, 
\quad B_{3}=-\frac{{\cal E}-\delta}{4}+\gamma(2N+\delta+1) ,\end{array} \quad
 %
  i\omega=-\gamma,\quad i\eta=N+\frac{\delta+1}{2}.
 \end{eqnarray}
 %
%
%
These parameters  lead to
 \begin{eqnarray}
 \begin{array}{l}
 \gamma_n^{(1)}=-2\gamma\left(n+N+\delta\right),\quad
\gamma_n^{(2)}=-2\gamma\left(n+N+\frac{1}{2}\right)
,\quad
\gamma_n^{(3)}=-2\gamma\left(n+N+1\right),\vspace{2mm}\\
\gamma_n^{(4)}=-2\gamma\left(n+N+\delta+\frac{1}{2}\right)
,\quad
 \gamma_n^{(5)}=2\gamma\left(n-N-1\right), \quad
\gamma_n^{(6)}=2\gamma\left(n-N-\delta-\frac{1}{2}\right),\vspace{2mm}\\
\gamma_n^{(7)}=2\gamma\left(n-N-\delta-1\right),\quad
\gamma_n^{(8)}=2\gamma\left(n-N-\frac{1}{2}\right),
\end{array}
 \end{eqnarray}
which must be used to construct the solutions.

\subsection{Solutions with finite series when $\ell$ is an integer}
%
 %
 %
 %
 Now we consider some solutions including integer values for $\ell$. From  $U^{(2)}(z)$, 
%
\begin{eqnarray}
\begin{cases}
\psi_{\;\ell}^{(2)}(z)=e^{-\gamma z}z^{\frac{3}{4}-\beta-\delta-\ell}
(z-1)^{\delta-\frac{1}{4}}
\displaystyle \sum_{n}b_{n}^{(2)}
z^{n},\qquad  \begin{cases}\mbox{if  } \ell=0,1,2,\cdots, \quad 0\leq n \leq \ell,\vspace{2mm}  \\
\mbox{otherwise,}\quad 0\leq n< \infty,
\end{cases}  \vspace{2mm}\\
\alpha_n^{(2)}=-\left[n+2-2(\beta+\delta+\ell)\right]
\left(n+1\right), \qquad 
\beta_n^{(2)}=
n\left(n+1+2\beta-2\ell\right)-\frac{1}{4}{\cal  E}- \frac{1}{8}\vspace{2mm}\\
-2(\gamma+\delta)(\beta+\delta+\ell),
\qquad 
\gamma_n^{(2)}=
-2\gamma\left(n-\ell-1\right).
\end{cases}\\
\begin{array}{l}\left[
\delta\geq\frac{1}{4}, \quad \beta+\delta+\ell\leq\frac{3}{4}:\quad\gamma<0\quad \mbox{for finite series},\quad
\quad \gamma\neq0 \quad \mbox{for infinite series}
\right]\end{array}. \nonumber 
\end{eqnarray}
Finite series for $\gamma>0$ are given by $\bm{\psi}_{\;\ell}^{(2)}$. 
Infinite series for $\ell=0,1,2,\cdots,$ come  from $U^{(6)}$ as implied by
\begin{eqnarray}
\psi_{\;\ell}^{(6)}(z)=
e^{\gamma z}z^{\frac{3}{4}-\beta-\delta-\ell}
(z-1)^{\delta-\frac{1}{4}}
\displaystyle \sum_{n=0}^{\infty}b_{n}^{(6)}
z^{n},\quad 
\delta\geq\frac{1}{4}, \quad \begin{array}{l}\beta+\delta+\ell\leq\frac{3}{4},\quad \ell=0,1,2,\cdots,\end{array}
\end{eqnarray}
where $\gamma_n^{(6)}=
2\gamma(n-2\beta-\ell+1)\neq 0$ for the conditions given above.


If $\ell \neq-1$  is a negative integer, finite series result from $(U^{(8)}, \bm{U}^{(8)})$. Thus,
\begin{eqnarray}
\begin{cases}
\psi_{\;\ell}^{(8)}(z)=  e^{\gamma z}z^{\beta+\delta+\ell-\frac{1}{4}}(z-1)^{\frac{3}{4}-\delta}
\displaystyle \sum_{n}b_{n}^{(8)}
z^{n},\qquad  \begin{cases}\mbox{if  } \ell=-2,-3.-4\cdots,\\
 \hspace{1.2cm} 0\leq n \leq -\ell-2,\vspace{2mm}  \\
\mbox{otherwise,}\quad 0\leq n< \infty,
\end{cases}  
\vspace{2mm}\\
%
\alpha_n^{(8)}=- \left[n+2(\beta+\delta+\ell)\right]
\left(n+1\right), \qquad \beta_n^{(8)}=
n\left(n+1+2\beta+2\ell)\right)-\frac{1}{4}{\cal E}-\frac{1}{8}\\
%
+2(\delta-1)(\beta+\ell+\delta),\qquad \gamma_n^{(8)}=
2\gamma \left(n+\ell+1\right).
\end{cases}\\
\begin{array}{l}\left[
\delta\leq\frac{3}{4}, \quad \beta+\delta+\ell\geq\frac{1}{4}:\quad 
\gamma>0 \quad \mbox{for finite series},
\quad \gamma\neq 0 \quad \mbox{for infinite series}
\right]\end{array}. \nonumber
\end{eqnarray}
Finite series for $\gamma< 0$ are given by $\bm{\psi}_{\;\ell}^{(8)}$.
Infinite series for $\ell=-2,-3,-4,\cdots,$ result from $U^{(4)}$,
\begin{eqnarray}
\psi_{\;\ell}^{(4)}(z)=e^{-\gamma z}z^{\beta+\delta+\ell-\frac{1}{4}}(z-1)^{\frac{3}{4}-\delta}
\displaystyle \sum_{n=0}^{\infty}b_{n}^{(4)}
z^{n},\quad \begin{array}{l}\delta\leq\frac{3}{4}, \quad \beta+\delta+\ell\geq\frac{1}{4},
\quad \ell=-2,-3,\cdots,\end{array}
%
\end{eqnarray}
where, under the above conditions, $\gamma_n^{(4)}=-2\gamma(n+2\beta+\ell-1)\neq 0$.

\section{Remarks on the  hyperbolic  potential given in  Eq.  (\ref{hiperbolico-II})}

For the Xie  potential (\ref{hiperbolico-II}), 
the substitutions
 \begin{equation}\label{xie-1}
z=\tanh^2u,\qquad \psi= (z-1)^{\frac{1}{2}\tau}\,U(z),\qquad \tau= \sqrt{-{\cal E}}
\end{equation}
transform the Schr\"{o}dinger equation (\ref{schr}) in
 %
 %
\begin{eqnarray}
\begin{array}{l}
z(z-1)\frac{d^{2}U}{dz^{2}}+
\left[-\frac{1}{2}+\left(\frac{3}{2}+\tau\right)z\right]
\frac{dU}{dz}+
\left[\frac{1}{4}\tau(\tau+1)
-\frac{v_3}{4}+\frac{v_1+v_2}{4}(z-1)-\frac{v_1}{4}z(z-1)\right]U=0,
\end{array}
\end{eqnarray}
 which is a CHE or a RCHE if $v_1\neq 0$ or $v_1=0$, respectively, with parameters 
 ($z_0=1$) 
 \begin{eqnarray}\label{hyper-a}
\begin{array}{l}
\mbox{CHE } (v_1\neq 0):\quad
B_{1}=-\frac{1}{2}, \quad  B_{2}=\frac{3}{2}+\tau, \quad 
 B_{3}=\frac{\tau(\tau+1)-v_3}{4},\quad
i\omega=\frac{\sqrt{v_1}}{2},\quad i\eta= \frac{v_1+v_2}{4\sqrt{v_1}};\vspace{3mm}\\
 %
 \mbox{RCHE } (v_1= 0): \quad B_{1}=-\frac{1}{2}, \quad  B_{2}=\frac{3}{2}+\tau, \quad 
 B_{3}=\frac{\tau(\tau+1)-v_3}{4},\quad
q=\frac{v_2}{4},\qquad U(z)=Y(z).
 \end{array}
 \end{eqnarray}
Since $(z-z_{0})^{1-B_{2}-\frac{B_{1}}{z_{0}}}=(z-1)^{=\tau}$, solutions bounded for
$z=1$ are obtained from the following expansions:
 \begin{eqnarray}
\begin{array}{l}
RCHE:\quad Y^{(1)},\; \qquad Y^{(2)}, 
\qquad Y(z)=U(z)\quad \mbox{  in Eq. (\ref{xie-1})};
 \vspace{2mm}\\
 %
CHE: \quad 
\left(U^{(1)}, \bm{U}^{(1)}\right), \quad \left(U^{(2)}, \bm{U}^{(2)}\right), \quad 
\left(U^{(5)}, \bm{U}^{(5)}\right), \quad \left(U^{(6)}, \bm{U}^{(6)}\right).
 \end{array}
 \end{eqnarray}

\subsection{Solutions for  $v_1=0$ (RCHE)}
The eigenfunctions are given by the infinite series
\begin{eqnarray}
\begin{cases}
\psi^{(1)}(u)=(z-1)^{\frac{!}{2}\tau}\displaystyle \sum_{n=0}^{\infty}b_{n}^{(1)}
z^{n},\qquad z=\operatorname{sech}^2 u;\\
\alpha_n^{(1)}=- \left(n+\frac{1}{2}\right)
\left(n+1\right), \qquad \beta_n^{(1)}=
n\left(n+\tau+\frac{1}{2}\right)+\frac{\tau(\tau+1)-v_2-v_3}{4},\qquad \gamma_n^{(1)}=\frac{v_2}{4};
\end{cases}
\end{eqnarray}
\begin{eqnarray}
\begin{cases}
\psi^{(2)}(u)=(z-1)^{\frac{1}{2}\tau} z^{\frac{1}{2}}\displaystyle \sum_{n=0}^{\infty}b_{n}^{(2)}
z^{n},\vspace{2mm}\\
\alpha_n^{(2)}=- \left(n+\frac{3}{2}\right)
\left(n+1\right), \quad \beta_n^{(2)}=
n\left(n+\tau+\frac{3}{2}\right)+\frac{\tau(\tau+1)-v_2-v_3}{4}+\frac{1}{2}(\tau+1),
\quad \gamma_n^{(2)}=\frac{v_2}{4}.
\end{cases}
\end{eqnarray}
The characteristic equations determine the relations among the parameters
$\tau$, $v_2$ and $ v_3$. The above solutions converge for  $0\leq z\leq 1$
as a consequence of (\ref{conve-1}).

Other problems satisfying the RCHE with 
$0\leq z\leq 1$ are given in \cite{lea-2,levy,dong}.
Incidentally, the equations of Refs. \cite{levy,dong}
deal with angular dependence the Schr\"{o}dinger 
equation for interaction of charged particles with electric dipoles.
These problems can be manage by means of the expansions of Sec. 2.6, or by 
series of special functions, as Bessel functions \cite{liu,lea-2}.


\subsection{Solutions for  $v_1\neq 0$ (CHE)}
 To examine the series, we write 
 \begin{eqnarray*}\begin{array}{ll}
  \gamma_n^{(1)}=\bm\gamma_n^{(1)}=
\sqrt{v_1}\left(n +\frac{v_1+v_2}{4\sqrt{v_1}} +\frac{\tau}{2}-\frac{1}{4}\right),\qquad
&
\gamma_n^{(2)}=\bm\gamma_n^{(2)}=
\sqrt{v_1}\left(n +\frac{v_1+v_2}{4\sqrt{v_1}} +\frac{\tau}{2}+\frac{1}{4}\right) , 
\vspace{2mm} \\
 \gamma_n^{(5)}=\bm\gamma_n^{(5)}=
-\sqrt{v_1}\left(n -\frac{v_1+v_2}{4\sqrt{v_1}} +\frac{\tau}{2}-\frac{1}{4}\right),\qquad
&
\gamma_n^{(6)}=\bm\gamma_n^{(6)}=
 -\sqrt{v_1}\left(n -\frac{v_1+v_2}{4\sqrt{v_1}} +\frac{\tau}{2}+\frac{1}{4}\right),$$
 \end{array}
 \end{eqnarray*}
Let us regard only the particular case $v_1=1$, choosen by Xie, 
 %
%
 %
 \begin{eqnarray}\begin{array}{ll}
  \gamma_n^{(1)}=\bm\gamma_n^{(1)}=
\left(n +\frac{v_2}{4} +\frac{\tau}{2}\right),\qquad
&
\gamma_n^{(2)}=\bm\gamma_n^{(2)}=
\left(n +\frac{v_2}{4} +\frac{\tau}{2}+\frac{1}{2}\right) , 
\vspace{2mm} \\
 \gamma_n^{(5)}=\bm\gamma_n^{(5)}=
-\left(n -\frac{v_2}{4} +\frac{\tau}{2}-\frac{1}{2}\right),\qquad
&
\gamma_n^{(6)}=\bm\gamma_n^{(6)}=
 -\left(n -\frac{v_2}{4} +\frac{\tau}{2}\right).$$
 \end{array}
 \end{eqnarray}
Thus,  if $v_2<0$, finite series can result from $
 \gamma_n^{(1)}$ and  $
 \gamma_n^{(2)}$;  if $v_2>0$,  finite series result from $
 \gamma_n^{(5)}$ and  $
 \gamma_n^{(6)}$. For example,
\begin{eqnarray}
\begin{cases}
\psi^{(1)}(u)=e^{\frac{1}{2} z}\;(z-1)^{\frac{1}{2}\tau}\displaystyle \sum_{n}b_{n}^{(1)}
z^{n},\qquad  \begin{cases}\mbox{if  } \frac{v_2}{4} +\frac{\tau}{2}=-N-1 , \quad 0\leq n \leq N,\vspace{2mm}  \\
\mbox{otherwise,}\quad 0\leq n< \infty,
\end{cases}\\
\alpha_n^{(1)}=- \left(n+\frac{1}{2}\right)
\left(n+1\right), \qquad \beta_n^{(1)}=
n\left(n+\tau-\frac{1}{2}\right)+\frac{1}{4}(\tau^2+\tau-v_2-v_3-2),\vspace{2mm}\\
 \gamma_n^{(1)}=\left(n +\frac{v_2}{4} +\frac{\tau}{2}\right),
\end{cases}
\end{eqnarray}
\begin{eqnarray}
\begin{cases}
\bm{\psi}^{(1)}(z)=e^{\frac{1}{2} z}\;(z-1)^{\frac{1}{2}\tau}\displaystyle \sum_{n}\bm{b}_{n}^{(1)}
(z-z_{0})^{n},\qquad  \begin{cases}\mbox{if  } \frac{v_2}{4} +\frac{\tau}{2}=-N-1 , \quad 0\leq n \leq N,\vspace{2mm}  \\
\mbox{otherwise,}\quad 0\leq n< \infty,
\end{cases}
\vspace{2mm}\\
\bm\alpha_n^{(1)}= \left(n+\tau+1\right)
\left(n+1\right),\quad \bm\beta_n^{(1)}=
n\left(n+\tau+\frac{3}{2}\right)+\frac{1}{4}(\tau^2+3\tau-v_3+2),\vspace{2mm}\\
\bm\gamma_n^{(1)}= \gamma_n^{(1)}=\left(n +\frac{v_2}{4} +\frac{\tau}{2}\right). \vspace{2mm}\\
\end{cases}
\end{eqnarray}
For $ \frac{v_2}{4} +\frac{\tau}{2}=-N-1 $, there are also the infinite series
  \begin{eqnarray}
\begin{cases}
\psi^{(5)}(u)=\bm\psi^{(5)}(u)=e^{-\frac{1}{2} z}\;(z-1)^{\frac{1}{2}\tau}\;\displaystyle \sum_{n=0}^{\infty}b_{n}^{(5)}
z^{n},\qquad    \begin{array}{l}    \frac{v_2}{4} +\frac{\tau}{2}=-N-1  \end{array} \vspace{2mm} \\
\alpha_n^{(5)}=- \left(n+\frac{1}{2}\right)
\left(n+1\right), \qquad \beta_n^{(5)}=
n\left(n+\tau+\frac{3}{2}\right)+\frac{1}{4}\left(\tau^2+\tau-v_2-v_3\right),\vspace{2mm}\\
 \gamma_n^{(5)}=-\left(n+N+ \tau+\frac{1}{2}\right).
\end{cases}
\end{eqnarray}
%


For the one-term series ($N=0$),  $\tau=-2-(v_2/2)$ while the characteristic equation $\beta_0^{(1)}=0$ 
(or $\bm\beta_0^{(1)}=0$) gives  the  relation among the parameters $v_i$:
\begin{eqnarray}
\left(\frac{v_2}{2}\right)^2+\frac{v_2}{2}-v_3=0, \qquad 
\psi^{(1)}_{N=0}(u)=\bm\psi^{(1)}_{N=0}(u)=b_0^{(1)}e^{\frac{1}{2} z}\;(z-1)^{\frac{1}{2}\tau}, 
\qquad[z=\tanh^2u].
\end{eqnarray}
Xie has exhibited only one-term series but we can get more terms by satisfying the 
characteristic equations. 
%
For two-term series ($N=1$),  we take $\tau=-4-(v_2/2)$ and then
$v_2/2$ is determined from $v_3$ by means of
\begin{eqnarray*}
&&\left(\frac{v_2}{2}\right)^4+8\left(\frac{v_2}{2}\right)^3+\left(21-2v_3\right)\left(\frac{v_2}{2}\right)^2+
\left(10-8v_3\right)\left(\frac{v_2}{2}\right)+\left[(v_3)^2-6v_3-48\right]
=0 \quad \mbox{for  } {\psi}^{(1)},\vspace{3mm}\\
&&\left(\frac{v_2}{2}\right)^4+6\left(\frac{v_2}{2}\right)^3+\left(11-2{v_3}\right)\left(\frac{v_2}{2}\right)^2-\left(10+6v_3\right)\left(\frac{v_2}{2}\right)+\left[(v_3)^2-6v_3-48\right]=0 \quad \mbox{for  } {\bm\psi}^{(1)}.
\end{eqnarray*}
%

The finite series with two or more terms do not satisfy the Arscott conditions (\ref{autovalores})  because we cannot write $\beta_i^{(j)}=\mathcal{B}_i^{(j)}-\Lambda$ for 
some constant $\Lambda$ which does not appear in  $\alpha_i^{(j)}$, $\mathcal{B}_i^{(j)}$
and $\gamma_i^{(j)}$.

Notice that in Ref. \cite{erro}  there is a particular case of the potential (\ref{hiperbolico-II}): 
 $v_2=v_3=0$.

  %
 
\section{Remarks on the hyperbolic  potential given in   Eq.  (\ref{hiperbolico-I})}

The Downing potential  (\ref{hiperbolico-I})  also appears in Ref.  \cite{erro}. By the substitutions
\begin{eqnarray}\label{psi}
z=\operatorname{sech}^2u, \qquad
\psi = z^{\tau/2}\;U,\qquad \tau=\sqrt{-{\cal E}},  \qquad\left[0\leq z\leq 1\right] 
\end{eqnarray}
the Schr\"{o}dinger 
equation is transformed into the CHE     
\begin{eqnarray}
z(z-1)\frac{d^{2}U}{dz^{2}}+
\left[\left(\frac{3}{2}+\tau\right)z- \tau-1\right]
\frac{dU}{dz}+
%
\frac{1}{4}\left[\tau(\tau+1)+{a}(z-1)-{a}z(z-1)\right]U=0,
\end{eqnarray}
with
 \begin{eqnarray}\label{hyper-1}
&&\begin{array}{l}
z_0=1,\quad B_{1}=-1-\tau, \quad  B_{2}=\frac{3}{2}+\tau, \quad 
  B_{3}=\frac{\tau(\tau+1)}{4},
 \quad
i\omega=-\frac{\sqrt{a}}{2},\quad i\eta=- \frac{\sqrt{a}}{4}.
 \end{array}
 \end{eqnarray}
Since $z^{1+\frac{B_1}{z_0}}=z^{-\tau}$, solutions bounded for
$z=0$ result only from the expansions:
 \begin{eqnarray*}
\begin{array}{l}
\left(U^{(1)}, \bm{U}^{(1)}\right), \quad \left(U^{(4)}, \bm{U}^{(4)}\right), \quad 
\left(U^{(5)}, \bm{U}^{(5)}\right), \quad \left(U^{(8)}, \bm{U}^{(8)}\right).
 \end{array}
 \end{eqnarray*}
In addition, the parameters (\ref{hyper-1}) imply that the coefficients 
$\gamma_n^{(1)}= \bm\gamma_n^{(1)}$ are
\begin{eqnarray}\label{gamas-3}
 \begin{array}{ll}											
 \gamma_n^{(1)}= \bm\gamma_n^{(1)}=-\sqrt{a}\left(n- \frac{\sqrt{a}}{4}+\frac{\tau}{2}-\frac{1}{4}\right),\quad
& \gamma_n^{(4)}=\bm\gamma_n^{(4)}=-\sqrt{a}\left(n - \frac{\sqrt{a}}{4}+\frac{\tau}{2}+\frac{1}{4}\right),\vspace{2mm}\\
\gamma_n^{(5)}=\bm\gamma_n^{(5)}=\sqrt{a}\left(n+\frac{\sqrt{a}}{4}+\frac{\tau}{2}-\frac{1}{4}\right),\quad&
 \gamma_n^{(8)}=\bm\gamma_n^{(8)}=\sqrt{a}\left(n + \frac{\sqrt{a}}{4}+\frac{\tau}{2}+\frac{1}{4}\right).
\end{array}
\end{eqnarray}
Hence, finite series require that 	
$\tau$ and $\sqrt{a}$ are such that $\gamma_n^{(i)}= \bm\gamma_n^{(i)}=0$ for some
$n\geq 1$; on the contrary, the expansions are given by infinite series. In any case, the characteristic equations give additional relations between $\tau$ and $\sqrt{a}$.

		
%
%
As an example, we consider $\left(U^{(1)}, \bm{U}^{(1)}\right)$.  
Finite series occur if $-\frac{\sqrt{a}}{4}+\frac{\tau}{2}=-N-\frac{3}{4}$; the
eigenfunctions (\ref{psi}) take the form
\begin{eqnarray}
\begin{cases}
\psi^{(1)}(u)=e^{-\frac{1}{2}\sqrt{a}\;z}\;z^{\frac{1}{2}\tau}\;\displaystyle \sum_{n}
b_{n}^{(1)}
z^{n},\qquad  \begin{cases}\mbox{if  } -\frac{\sqrt{a}}{4}+\frac{\tau}{2}=-N-\frac{3}{4} , \quad 0\leq n \leq N,\vspace{2mm}  \\
\mbox{otherwise,}\quad 0\leq n< \infty,
\end{cases}
\vspace{2mm}\\
\alpha_n^{(1)}=- \left(n+\tau+1\right)
\left(n+1\right), 
\quad
 \beta_n^{(1)}=
n\left(n+\tau+\sqrt{a}+\frac{1}{2}\right)+ 
\frac{\tau(\tau+1)}{4}
-\frac{a}{4} + 
 \frac{\sqrt{a}}{2}(1+\tau),
\vspace{2mm}\\
 \gamma_n^{(1)}=
-\sqrt{a}\left(n- \frac{\sqrt{a}}{4}+\frac{\tau}{2}-\frac{1}{4}\right);
\end{cases}
\end{eqnarray}
%
%
						%
\begin{eqnarray}
\begin{cases}
\bm\psi^{(1)}(u)=e^{-\frac{1}{2}\sqrt{a}\;z}\;z^{\frac{1}{2}\tau}\;\displaystyle \sum_{n}
\bm{b}_{n}^{(2)}
(z-1)^{n},\qquad  \begin{cases}\mbox{if  } -\frac{\sqrt{a}}{4}+\frac{\tau}{2}=-N-\frac{3}{4} , \quad 0\leq n \leq N,\vspace{2mm}  \\
\mbox{otherwise,}\quad 0\leq n< \infty,
\end{cases}
\vspace{2mm}\\
\bm\alpha_n^{(1)}=\left(n+\frac{!}{2}\right)
\left(n+1\right), 
\qquad
\bm \beta_n^{(1)}=
n\left(n+\tau-\sqrt{a}+\frac{1}{2}\right)+ \frac{\tau(\tau+1)}{4}-\frac{\sqrt{a}}{4},
\qquad \bm\gamma_n^{(1)}=\gamma_n^{(1)}.
\end{cases}
\end{eqnarray}
%
%
Since
%
\begin{eqnarray*}\begin{array}{l}
-\frac{\sqrt{a}}{4}+\frac{\tau}{2}=-N-\frac{3}{4} \;\Rightarrow\;
\gamma_n^{(1)}=\bm\gamma_n^{(1)}=
-\sqrt{a}\left(n- N-1\right) \mbox{ and } \gamma_n^{(5)}=\bm\gamma_n^{(5)}=
\sqrt{a}\left(n+ N+\tau+\frac{1}{2}\right),
\end{array}
\end{eqnarray*}
%
in this case, there is an expansion in infinite series represented by
  \begin{eqnarray}
\begin{cases}
\psi^{(5)}(u)=\bm\psi^{(5)}(u)=e^{\frac{1}{2}\sqrt{a}\;z}\;z^{\frac{1}{2}\tau}\;\displaystyle \sum_{n=0}^{\infty}b_{n}^{(5)}
z^{n},\qquad    \begin{array}{l} -\frac{\sqrt{a}}{4}+\frac{\tau}{2}=-N-\frac{3}{4},\end{array} \vspace{2mm} \\
\alpha_n^{(5)}=- \left(n+\tau+1\right)
\left(n+1\right), \qquad \beta_n^{(5)}=
n\left(n+\tau-\sqrt{a}+\frac{1}{2}\right)+\frac{\tau(\tau+1)}{4}-\frac{a}{4}
-\frac{\sqrt{a}}{4}(\tau+1),
\vspace{2mm}\\
 \gamma_n^{(5)}=
\sqrt{a}\left(n+N+ \tau+\frac{1}{2}\right).
\end{cases}
\end{eqnarray}
%
%
Therefore, there are infinite series if $-\frac{\sqrt{a}}{4}+\frac{\tau}{2}=-N-\frac{3}{4}$
and if $-\frac{\sqrt{a}}{4}+\frac{\tau}{2}\neq-N-\frac{3}{4}$.


For the one-term series ($N=0$),  $\sqrt{a}=2\tau+3$ while the characteristic equation $\beta_0^{(1)}=0$ 
(or $\bm\beta_0^{(1)}=0$) implies that $\tau^2-\tau-3=0$. Since $\tau>0$, we obtain
\begin{eqnarray}
\psi^{(1)}(u)=\bm\psi^{(1)}(u)=b_0^{(1)}e^{-\sqrt{a}\;z/2}\;z^{\tau/2},\quad
\sqrt{a}=4+\sqrt{13},\quad \tau=[1+\sqrt{13}]/2.
\end{eqnarray}
%
%
For the two-term series ($N=1$), the condition $\sqrt{a}=2\tau+7$  and the  
characteristic equations $\beta_{0}\beta_{1}- \alpha_{0}\gamma_{1} = 0$ give
\begin{eqnarray}
\begin{array}{ll}
\tau^4-6\tau^3-31\tau^2+48\tau+147=0,\quad &[\sqrt{a}=2\tau+7],
%
\end{array}
\end{eqnarray}
for both expansions, $\psi^{(1)}$ and $\bm\psi^{(1)}$. This fact suggests degenerated  eigenfunctions.
%


%
%
%
	%
%
%
%
%
%
%
%
%

%
											
%

\section{Conclusions}

Two noteworthy points of the present analysis are the region of convergence of the power series solutions and the use of the transfomations (\ref{trans}).  These imply  infinite-series solutions
for the same set of parameters which provides finite-series solutions.

In effect, before using the power series  solutions of the CHE,  in Sec. 2.4 we have seen that
these expansions  converge for finite values of $z$, including the regular singular points $z=0$ and
$z=z_0$. 
The  region $|z|<|z_0|$,  
obtained from the expression (\ref{dominant}) for the ratio test, cannot be used because the corresponding characteristic  equation does not converge. Thus, we have enlarged the misleading region $|z|<|z_0|$ which appears in a book on  Heun equations \cite{ronveaux}.

On the other hand, the transformations (\ref{trans}) generate eight pairs of solutions $\left(U^{(i)},\bm{U}^{(i)}\right)$  for the
CHE,  where  $U^{(i)}$ and $\bm{U}^{(i)}$  are expansions in series of $z$  and $z-z_0$, respectively.  
 For the QES under consideration, the finite series occur only for special sets of the parameters; 
 however, infinite
series occur for all admissible values of the parameters, even for the values which give finite series. 
This follows from the fact summarized in (\ref{prescription}) as: if the pair $\left({U}^{(i)},\bm{U}^{(i)}\right)$ 
 gives finite series, then $\left({U}^{(i+4)},\bm{U}^{(i+4)}\right)$ gives infinite series; 
 if $\left({U}^{(i+1)},\bm{U}^{(i+1)}\right)$ 
 gives finite series, then $\left({U}^{(i)},\bm{U}^{(i)}\right)$ gives infinite series ($i=1,2,3,4$).
 Observe that, apart from the bounded  factors $\exp{(\pm i\omega z)}$,  the series are 
 multiplied by the same function in both pairs.
   
 As a matter of fact,  we have dealt with classes of potentials whose parameters
 must be specified. In Sec. 3.1 we have selected a particular case 
 of the potential  (\ref{ush-1-trigonometrico}), obeying  a 
Whittaker-Hill equation. 
For this case all the expansions  $(U^{(i)},\bm{U}^{(i)})$ afford finite-series 
eigenfunctions when $\ell$ is  integer or  half an odd integer, a fact that 
does not preclude  infinite series.
At the end, we have show how to apply the same procedure to the Razavy trigonometric potential 
(\ref{razavy-trigonometrico})  concerning  oscillations of a molecule.
For $\bm{n}=-1$, the Razavy potential provides a Mathieu equation  letting only 
infinite-series solutions.

For the trigonometric potential  (\ref{ush-2-trigonometrico}),  in Sect. 4.1 we have  displayed  
a case which admits finite-series solutions when $\ell$ is integer or half an odd integer, 
and also infinite-series solutions. 
At the end of the section, we have explained  how to adapt the solutions to the
trigonometric potential written in Eq. (\ref{pozo}).

The finite-series solutions for the trigonometric potentials, unlike the 
hyperbolic case,   
satisfy the Arscott conditions (\ref{autovalores})  that insure real and distinct 
eigenvalues of energy.  
In Secs. 5 and 6  we have seen that the eight pairs of solutions
may add new finite-series  solutions for the hyperbolic potentials and also 
infinite series.
If $v_1=0$ in the potential (\ref{hiperbolico-II}), the equation reduces to a RCHE  that admits only  infinite-series 
solutions, as seen in Sec.  5.1.

Although the power series solutions for the CHE are efficient for providing finite series, 
the complete solution for these QES potentials 
requires the computation of infinite series solutions,
not necessarily in terms of power series. 
Up to now only finite series  have been taken into account for QES problems. 
J. W. Liu  have  
discussed methods  
suitable for  treating infinite series of the angular CHE \cite{liu}. 

Finally we observe that for some QES 
potentials given by elliptic functions 
the Schr\"{o}dinger
  equation can be solved by using solutions of Heun's general equation: once more, the problem  admits  infinite-series solutions 
corresponding to finite-series solutions
  \cite{2021,2022}.

\section*{Acknowledgement}
 
%
This work was funded by Minist\'erio da Ci\^encia, Tecnologia e Inova\c{c}\~ oes do Brasil. 
The work does not have conflicts of interest. The author thanks 
Dr. Maria Santos Reis Bonorino Figueiredo for reading the manuscript.

\section*{ORCID}

Bartolomeu D. B. Figueiredo:  
orcid.org/0000-0001-6346-3991

%

%
%
%

%
\end{document}